\documentclass[structabstract,usenatbib,useAMS]{aa}  
\usepackage{graphicx}
\usepackage{txfonts}
\usepackage{natbib}
\usepackage{amssymb}
\newcommand{\ha}{\relax \ifmmode {\mbox H}\alpha\else H$\alpha$\fi}
\newcommand{\hb}{\relax \ifmmode {\mbox H}\beta\else H$\beta$\fi}
\newcommand{\sii}{\relax \ifmmode {\mbox S\,{\scshape ii}}\else S\,{\scshape ii}\fi}
\newcommand{\nii}{\relax \ifmmode {\mbox N\,{\scshape ii}}\else N\,{\scshape ii}\fi}
\newcommand{\oii}{\relax \ifmmode {\mbox O\,{\scshape ii}}\else O\,{\scshape ii}\fi}
\newcommand{\oiii}{\relax \ifmmode {\mbox O\,{\scshape iii}}\else O\,{\scshape iii}\fi}
\begin{document}
   \title{A Virtual Observatory Census to Address Dwarfs Origins}

   \subtitle{AVOCADO -- I. Science goals, sample selection, and analysis tools}

   \author{R. S\'anchez-Janssen
          \inst{1}
          \and
          R. Amor\'in
          \inst{2}
          \and
          M. Garc\'ia-Vargas
          \inst{3}
          \and
          J.M. Gomes
          \inst{4}
          \and
          M. Huertas-Company
          \inst{5}
          \and
          F. Jim\'enez-Esteban
          \inst{6,7,8}
          \and
          M. Moll\'a
          \inst{9}
          \and
          P. Papaderos
          \inst{4}
          \and
          E. P\'erez-Montero
          \inst{2}
          \and
          C. Rodrigo
          \inst{6,7}
          \and
          J. S\'anchez Almeida
          \inst{10,11}
          \and
          E. Solano
          \inst{6,7}
          }

   \institute{European Southern Observatory,
              Alonso de C\'ordova 3107, Vitacura, Santiago, Chile\\
              \email{rsanchez@eso.org}
              \and Instituto de Astrof\'isica de Andaluc\'ia-CSIC, Glorieta de la Astronom\'ia S/N, E-18008 Granada, Spain
              \and FRACTAL SLNE, C/Tulip\'an 2, p13 1-A, E-28231 Las Rozas (Madrid), Spain
              \and Centro de Astrof\'isica and Faculdade de Ci\^encias, Universidade do Porto, Rua das Estrelas, 4150-762, Porto, Portugal
              \and GEPI, Paris-Meudon Observatory, 5 Place Jules Janssen, 92190 Meudon, France
              \and Centro de Astrobiolog\'{\i}a (INTA-CSIC), Departamento de Astrof\'{\i}sica, PO Box 78, E-28691, Villanueva de la Ca\~nada, Madrid, Spain
	    \and Spanish Virtual Observatory, Spain
	    \and Saint Louis University, Madrid Campus, Division of Science and Engineering, Avenida~del~Valle 34, E-28003 Madrid, Spain
	    \and Departamento de Investigaci\'{o}n B\'{a}sica, CIEMAT, Avda. Complutense 40. E-28040 Madrid, Spain
	    \and Instituto de Astrof\'isica de Canarias, E-38205 La Laguna, Tenerife, Spain
	    \and Departamento de Astrof\'isica, Universidad de La Laguna, E- 38071 La Laguna, Tenerife, Spain
             }

   \date{Received; accepted}

% \abstract{}{}{}{}{} 
% 5 {} token are mandatory
 
  \abstract
 {Even though they are by far the most abundant of all galaxy types, the detailed properties of dwarf galaxies are still only poorly characterised -- especially because of the observational challenge that their intrinsic faintness and weak clustering properties represent.}
 {AVOCADO aims at establishing firm conclusions on the formation and evolution of dwarf galaxies by constructing and analysing a homogeneous, multiwavelength dataset for a statistically significant sample of approximately 6500 nearby dwarfs ($M_{i}-5\,\mbox{log}\,h_{100} > -18$ mag). The sample is selected to lie within the $20 < \mbox{D} < 60~h_{100}^{-1}$ Mpc volume covered by the SDSS-DR7 footprint, and is thus volume-limited for $M_{i}-5\,\mbox{log}\,h_{100}<-16$ mag dwarfs -- but includes $\approx$\,1500 fainter systems. We will investigate the roles of mass and environment in determining the current properties of the different dwarf morphological types -- including their structure, their star formation activity, their chemical enrichment history, and a breakdown of their stellar, dust, and gas content.}
 {We present the sample selection criteria and describe the suite of analysis tools, some of them developed in the framework of the Virtual Observatory. We use optical spectra and UV-to-NIR imaging of the dwarf sample  to derive star formation rates, stellar masses, ages, and metallicities -- which are supplemented with structural parameters that are used to classify them morphologically. This unique dataset, coupled with a detailed characterisation of each dwarf's environment, allows for a fully comprehensive investigation of their origins and enables us to track the (potential) evolutionary paths between the different dwarf types.} 
 {We characterise the local environment of all dwarfs in our sample, paying special attention to trends with current star formation activity. 
We find that virtually all quiescent dwarfs are located in the vicinity (projected distances $\lesssim$\,$1.5\,h_{100}^{-1}$ Mpc) of $\gtrsim L^{*}$ companions, consistent with recent results. 
While star-forming dwarfs are preferentially found at separations of the order of 1\,$h_{100}^{-1}$ Mpc, there appears to be a tail towards low separations ($\lesssim$\,$100\,h_{100}^{-1}$ kpc) in the distribution of projected distances. We speculate that, modulo projection effects, this probably represents a genuine population of late-type dwarfs caught upon first infall about their host and before environmental quenching has fully operated.
In this context, these results suggest that internal mechanisms --such as gas exhaustion via star formation or feedback effects-- are not sufficient to completely cease the star formation activity in dwarf galaxies, and that becoming the satellite of a massive central galaxy appears to be a \emph{necessary} condition to create a quiescent dwarf.
}
   {}


   \maketitle
%
%________________________________________________________________

\section{Introduction}
Dwarf galaxies are the most abundant of all galaxy types in the Local Universe \citep{Blanton2005b}, and this simple fact makes them key objects to constrain galaxy formation and evolution models.
Notably, their high abundance is nothing but the manifestation of the underlying steep mass function of dark matter (DM) haloes \citep{Klypin1999}. 
Indeed, it is currently well established that the low-mass haloes where dwarf galaxies probably reside are extremely inefficient in retaining baryons and in converting them into stars, as indicated by their observed high total-to-baryonic \citep[][]{Strigari2008,Walker2009} and gas mass fractions \citep[][]{Geha2006,Warren2007}.

However, and despite this apparent inefficiency, the peak of the specific star formation rate (sSFR, or SFR per unit stellar mass) shifts towards lower masses at lower redshifts (e.g., \citealt{Martin2007}). This implies that lower mass galaxies have experienced more prolonged star formation relative to more massive ones -- another manifestation of the well-known downsizing effect \citep{Cowie1996}. This trend has been further confirmed from the detailed analysis of resolved stellar populations in nearby dwarfs: their star formation histories (SFHs) appear to deviate from the cosmic mean at $z \lesssim 0.7$ \citep{Weisz2011}. Indeed, the bulk of current stellar mass in most Local Volume dwarfs formed more than 7 Gyr ago, with differences between early- and late-type dwarfs only arising during the last 1 Gyr \citep{Weisz2011}. 
However, despite this apparent uniformity, SFHs are far from simple \citep{Grebel1997,Mateo1998,Dolphin2002,Tolstoy2009,Weisz2011}, differing from the most commonly assumed behaviours -- exponentially declining, constant, or single-burst SFRs. The actual duration of starbursts and their role in the SFHs of the different dwarf types are still a matter of intense debate \citep{Ostlin2001,SanchezAlmeida2008,Lee2009,McQuinn2009} 
This complexity in SFHs most probably reflects the extreme conditions in which star formation occurs in dwarf galaxies.
It is not for nothing that these low-mass systems have long been considered excellent laboratories for the study of star formation and feedback effects \citep[e.g., ][]{Hunter1998}. They have high sSFRs \citep{Lee2009a}, low metallicities \citep{Skillman1989,Izotov2006}, low dust content \citep{Hunter1989}, and preferentially inhabit shallow potential wells \citep{Simon2005,deBlok2008} where turbulent motions are as important as, and possibly even more than, angular momentum support \citep{Kaufmann2007,Schroyen2011}. All these conditions are significantly different from those of more massive galaxies, and most likely resemble those prevailing at high redshift \citep{Brammer2012}. 

In particular, the two latter characteristics -- low masses and low angular momentum support -- make dwarf galaxies very sensitive to the effects of both stellar feedback processes and environmental mechanisms. 
From the theoretical point of view, it is well established that strong feedback effects due to supernovae and intense star formation can regulate the chemical and structural properties of low-mass galaxies, as they can lead to the removal of low-angular momentum material -- be it dark matter of the interstellar medium \citep{Dekel1986,Ferrara2000,Dekel2003a,Governato2010} --, thus modifying their scaling relations. Feedback is therefore thought to be the main internal factor regulating the gas content, star formation activity, and metal production in dwarf galaxies -- as they mostly lack spiral arms \citep{Gallagher1984} and/or bars \citep{Mendez-Abreu2010,Mendez-Abreu2012}.
On the other hand, the well-known morphology-density relation of bright galaxies \citep{Dressler1980} extends towards lower luminosities: early-type dwarfs in the Local Universe tend to inhabit higher density environments than late-types \citep[e.g.,][]{Binggeli1991,Vilchez1995,Karachentsev2004,Weisz2011}. This strongly suggests that environmental effects must be responsible, to a significant degree, for the transformation from actively star forming to quiescent dwarfs \citep[e.g.,][]{Grebel2003}. The similarities between the structural properties and SFHs of these two families have long been used to argue for the existence of evolutionary paths between the different dwarf types \citep{Papaderos1996,vanZee2004,GildePaz2005}. Several caveats apply, however. 
For instance, \citet{Papaderos1996} show that evolutionary connections between dwarf ellipticals (dEs), blue compact dwarfs (BCDs), and dwarf irregulars (dIs) are only plausible if the BCD host galaxy can modify its structural properties -- possibly through adiabatic contraction and/or expansion in response to gas infall and/or outflows.
It is obvious that, at some point during their evolutionary history, early-type dwarfs lost and/or exhausted their gas reservoirs, and ceased forming stars -- possibly by a combination of both internal \citep{Dekel1986} and external factors \citep[e.g.,][]{Mayer2001a,Grebel2003}. One must keep in mind, however, that these evolutionary connections most likely only concern the \emph{ancestors} of current early- and late-type dwarfs \citep{Skillman1995,Ferrara2000,rsj2012}.

Despite their critical importance, dwarf galaxy properties are still only poorly understood -- both from the observational and theoretical  side.
Simulations cannot yet fully reproduce the process of dwarf formation, as they require challengingly high resolutions to accurately describe the complex baryon physics that dominate these small scales. Thus, star formation occurs too early and fast in semi-analytic models of  dwarf formation, resulting in an excess of faint red systems with respect to observations (e.g., \citealt{Henriques2008}). Moreover, all detailed hydrodynamical simulations produce dwarfs with too high baryonic masses for their corresponding dark matter haloes \citep{Sawala2011}. 

From an observational point of view, very detailed studies from resolved stellar populations have been necessarily restricted to the Local Volume. This significantly limits their statistical power, with samples rarely exceeding a few dozens and being subject to cosmic variance effects. This results in high incompleteness of the most extremes dwarf examples, including the most metal-poor systems \citep{Kunth2000}, the brightest starbursts, and those residing in very low and very high density environments. Beyond the Local Volume, the picture does not change significantly. The intrinsic low luminosities of dwarfs and their weak clustering properties have so far prevented studies of statistically significant field dwarf samples, with the most complete ones barely exceeding a few hundred objects (e.g., \citealt{GildePaz2003,Geha2006,Hunter2006,Lee2009}). This effect even worsens at higher redshifts ($z > 0.1$), where field dwarf galaxies are rather poorly characterised. Cluster dwarfs constitute an exception, with large samples allowing a better understanding of both low- \citep{rsj2008} and intermediate-redshift \citep{Barazza2009} cluster dwarfs.

AVOCADO (A Virtual Observatory Census to Address Dwarfs Origins) aims at providing strong constraints on dwarf galaxy formation and evolution by constructing an unprecedented, homogeneous, multiwavelength dataset for a statistically significant sample of several thousand nearby dwarfs. 
We  use  optical spectra and UV-to-NIR imaging of the dwarf sample  to derive star formation rates, stellar masses, ages, and metallicities -- which, when further supplemented with structural parameters and a detailed characterisation of each dwarf's environment, allow for a fully comprehensive investigation of their origins.
The paper is organised as follows. Section\,\ref{sect:sample} describes the sample selection, together with the imaging and spectroscopic datasets. Sections\,\ref{sect:VOSA} and \ref{sect:spectra} present a brief description of our adopted approach to analyse the UV-to-NIR imaging and the optical spectra. Section\,\ref{sect:morph} focuses on the determination of morphological types, while in Section\,\ref{sect:environment} we characterise the environment in which dwarfs reside. Finally, Section\,\ref{sect:gas} summarises the available information on the neutral and molecular gas content of AVOCADO dwarfs. The scientific results of the project will be presented in future papers of this series.

Throughout  this paper we assume a flat cosmology with $\Omega_{M} = 0.3$, $\Omega_{\Lambda} = 0.7$ and $H_{0} = 100$ km\,s$^{-1}$\,Mpc$^{-1}$.

%________________________________________________________________

\section{AVOCADO: sample and datasets}
\label{sect:sample}

\subsection{Definition of dwarf galaxy and sample selection}
The classification of a dwarf galaxy has long been recognised to be an ill-posed problem. The simplest criterium refers to a low-mass system, but the selection of a characteristic scale is not a simple task. 
Moreover, given that only a small fraction of a dwarf galaxy mass is in the form of stars, one has to wonder whether stellar, baryonic, or total masses should be considered. The two latter options are usually impractical, because total masses require kinematical information and are essentially limited to Local Group dwarfs \citep{Strigari2008,Walker2009}, while the neutral gas content has been missing for any statistically significant dwarf sample until very recently \citep{Geha2006,Begum2008a,Huang2012}.

The majority of galactic properties -- including stellar populations, structure and kinematics -- correlate with stellar mass (or luminosity), and it is well known that these relations show a high degree of continuity. Notably, however, the stellar Tully-Fisher relation exhibits a clear break at low masses, with V$_{c} < 90$ km\,s$^{-1}$ galaxies  falling below the relation defined by more massive objects -- but note that the break vanishes when \emph{baryonic} masses are instead considered \citep{McGaugh2000}. Moreover, this limit corresponds to the scale where galaxies start to be systematically thicker \citep{rsj2010}, deviating from the mass-size relation of disc galaxies \citep{Schombert2006} -- most likely a result of the increasing importance of turbulent motions over rotational support. 
Circular velocities are not ideal observables from a practical point of view, but they can be easily transformed to luminosities through the previously mentioned Tully-Fisher relation. According to the relation derived by \citet{Geha2006}, the limiting value of V$_{c} < 90$ km\,s$^{-1}$ corresponds to $M_{i}-5\,\mbox{log}\,h_{100} > -18$ mag, or $M^{*}+2.5$ \citep{Blanton2005b}. We therefore adopt this upper luminosity limit to define our dwarfs, guaranteeing that the AVOCADO sample genuinely contains low-mass systems.

\begin{figure}
\centering
\includegraphics[width=.5\textwidth,clip=true]{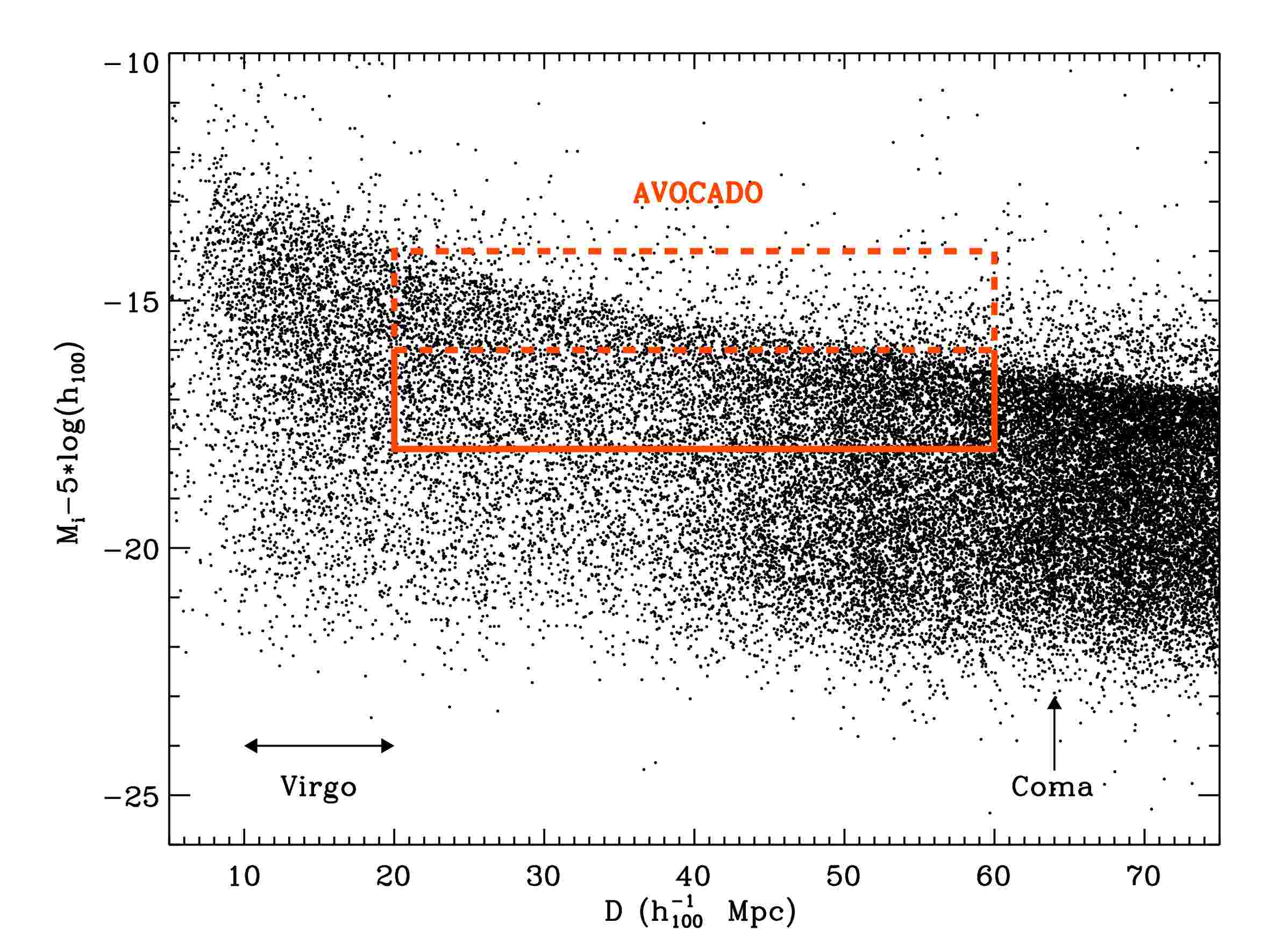} 
\caption{Magnitude-distance relation for nearby galaxies in the parent NASA-Sloan sample. The $\approx$\,6500 AVOCADO dwarfs have been selected to lie within distances $20 < \mbox{D} < 60~h_{100}^{-1}$ Mpc and be fainter than $M_{i}-5\,\mbox{log}\,h_{100} > -18$ mag (or $M^{*}+2.5$). This ensures that the sample is volume-limited for objects $M_{i}-5\,\mbox{log}\,h_{100}\lesssim -16$ mag, furthermore including $\approx$\,1500 fainter systems down to $M_{i}-5\,\mbox{log}\,h_{100} \approx -14$ mag. The (peculiar-velocity-corrected) distance range also excludes the bulk of the Virgo and Coma dwarf galaxy populations.} 
\label{fig:magdist}
\end{figure}

The goal of constructing a statistically significant sample of nearby dwarf galaxies naturally leads to the selection of large-area, multiwavelength public surveys.
Our parent sample is drawn from the NASA-Sloan Atlas \citep{Blanton2011} for nearby galaxies, an extension of their NYU-VAGC \citep{Blanton2005a}. For consistency with the analysis of spectra (cf. Section\,\ref{sect:spectra}), we only consider objects within the SDSS-DR7 footprint. Redshifts for $\approx$\,90\% of the AVOCADO sample were obtained from the SDSS spectra, or other sources when available. 
Including these additional redshifts increases the number of available candidates on scales $<55$ arcsec, where SDSS fibre collisions  produce incompleteness.
We recall that the SDSS spectroscopic target selection is essentially unbiased, with no prior selection on magnitude, colour, or surface brightness. It should therefore not introduce any non-uniformity in terms of any intrinsic galaxy property.

In order to select a sample of nearby galaxies, a (peculiar-velocity-corrected) distance cut $20 < \mbox{D} < 60~h_{100}^{-1}$ Mpc was introduced (see Fig.\,\ref{fig:magdist}). 
The lower limit was set to avoid strong relative corrections from peculiar motions, while the upper one ensures that the parent sample is volume-limited for objects $M_{i}-5\,\mbox{log}\,h_{100}\lesssim -16$ and a surface brightness $<$$\mu$$>$$_{r,50}< 23.5$ mag\,arcsec$^{-2}$ \citep{Blanton2005b}. 
Moreover, and as indicated in Fig.\,\ref{fig:magdist}, these distance boundaries exclude the bulk of the Virgo and Coma cluster galaxies, whose dwarf populations have already been studied with greater detail in the literature \citep[e.g.,][and many others]{Binggeli1985,Impey1988,Binggeli1991,Ferrarese2006,Lisker2007,Boselli2008,Secker1997,Trentham1998,Poggianti2001,Aguerri2005,Smith2009}.

\begin{figure}
\centering
\includegraphics[width=.5\textwidth,clip=true]{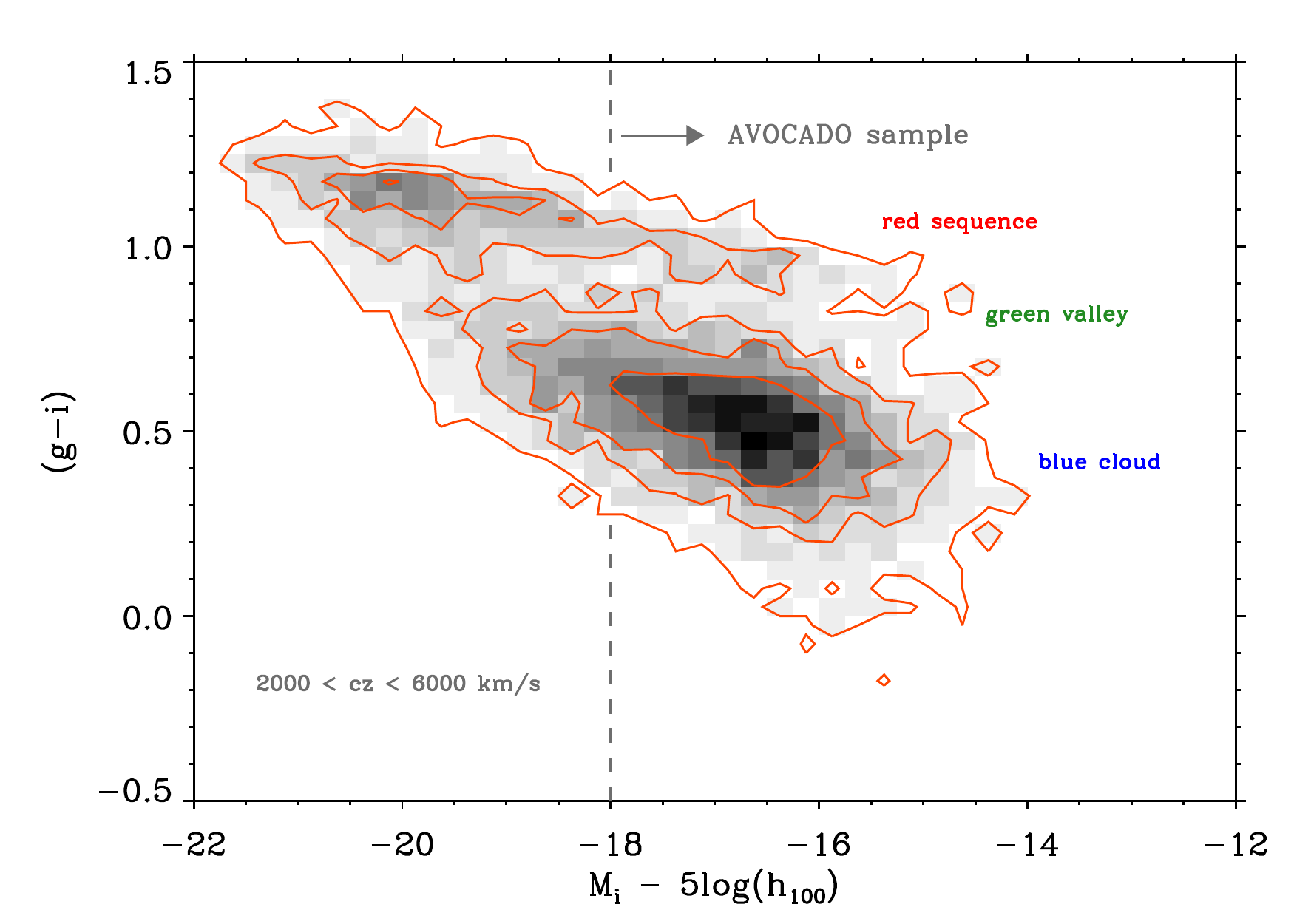} 
\caption{Colour-magnitude diagram (CMD) of the parent sample in the $20 < \mbox{D} < 60~h_{100}^{-1}$ Mpc distance range. The shaded region indicates the (not volume-corrected) number density distribution of galaxies in each CMD bin, while the contours enclose 95\%, 75\%, 50\%, and 25\% of the population. The dashed vertical line indicates our bright magnitude limit ($M_{i}-5\,\mbox{log}\,h_{100}=-18$) for the selection of the AVOCADO sample. Note that, even though no massive clusters are included, the red sequence, green valley, and blue cloud are clearly distinguishable in the CMD, the latter being by far the dominant population.} 
\label{fig:cmd}
\end{figure}

\begin{figure*}
\includegraphics[width=0.55\textwidth,clip=true]{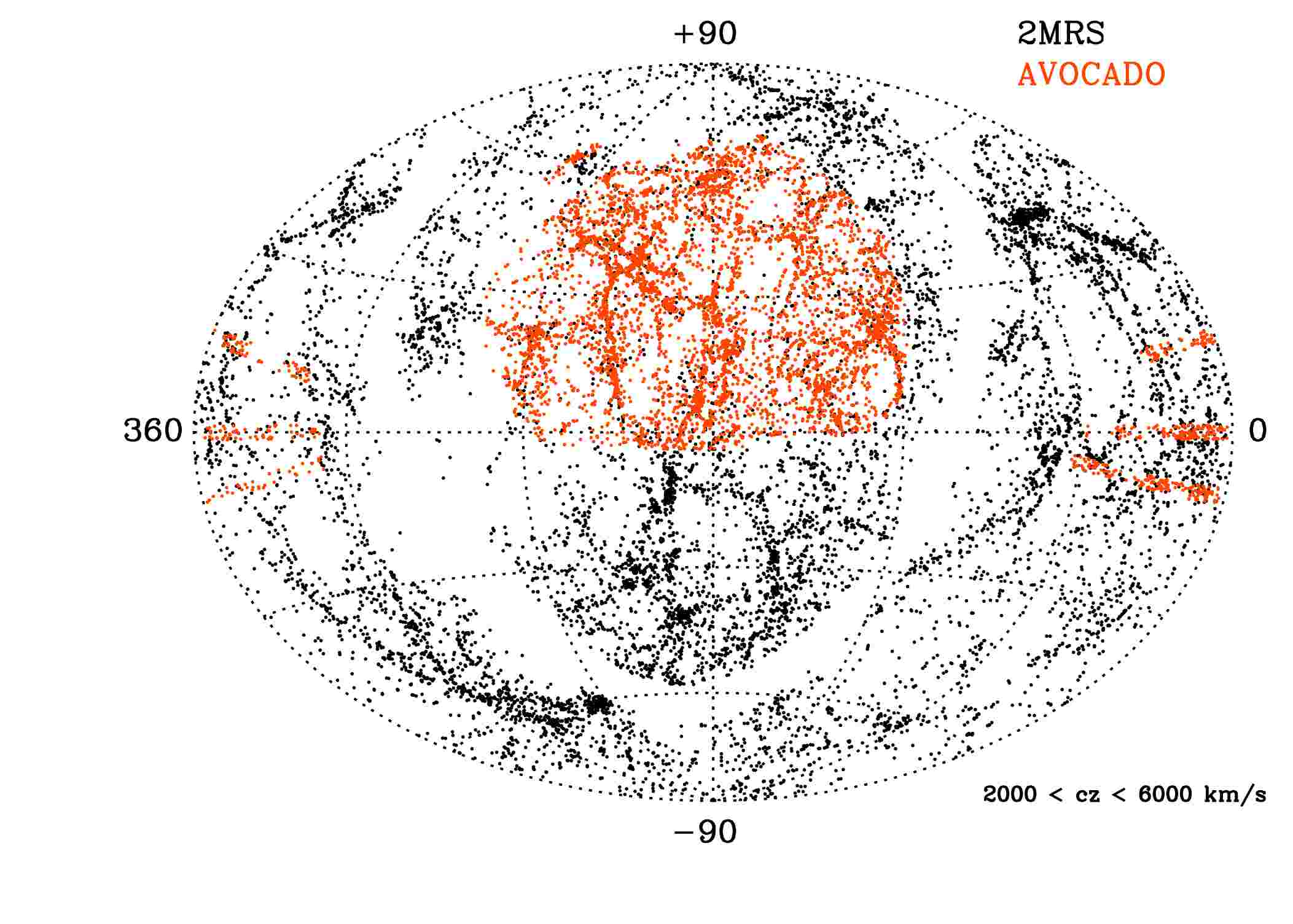}
\includegraphics[width=0.55\textwidth,clip=true]{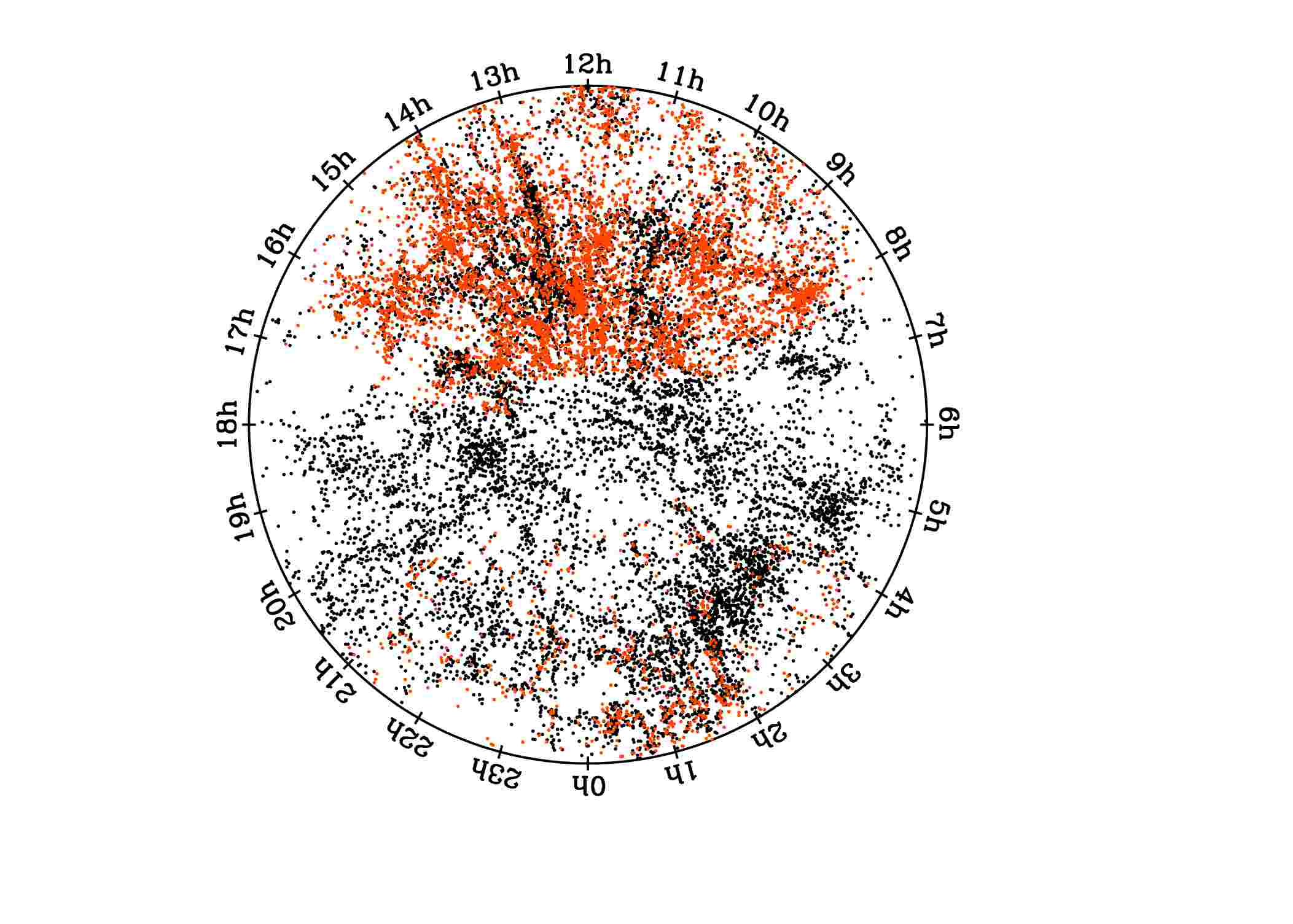}
\caption{
\emph{Left}: Aitoff projection of the spatial distribution of dwarfs in the AVOCADO sample (orange) and of $K < 11.75$\,mag galaxies in the 2MASS Redshift Survey (black, \citealt{Huchra2011}) in the $2000 < cz < 6000$ km\,s$^{-1}$ range. Note that dwarf galaxies perfectly trace the underlying large-scale structure of voids and filaments. The 2MRS, being an all-sky, magnitude-limited survey, provides an excellent sample for studing the environment in which AVOCADO dwarfs evolve.
\emph{Right}: Hockey-puck plot (see \citealt{Huchra2011}) of the same two samples, showing a projection from the northern celestial pole.
}
\label{fig:spatial}
\end{figure*}

Figure\,\ref{fig:cmd} shows the ($g-i$) colour-magnitude diagram of the parent sample, with the grey scale indicating the (not volume-corrected) number density distribution of galaxies. The upper luminosity limit adopted for the selection of the AVOCADO sample is indicated together with the locations of the characteristic red sequence, green valley, and blue cloud.
The final sample consists of 6588 dwarf galaxies, $\sim$1500 of which are fainter than $M_{i}-5\,\mbox{log}\,h_{100} = -16$. This sample size represents an order of magnitude increase with respect to any previous detailed spectroscopic and multiwavelength imaging studies of dwarf galaxies. 
Fig.\,\ref{fig:spatial} (left) shows an Aitoff projection of the sky distribution of AVOCADO dwarfs (orange) compared to the all-sky bright galaxy sample of the 2MASS Redshift Survey \citep[][2MRS hereafter]{Huchra2011} in the $2000 < cz < 6000$ km\,s$^{-1}$ range. Dwarf galaxies inhabit both low- and intermediate-density environments, nicely tracing the underlying large-scale structure of voids and filaments. This will allow us to explore the effects of halo mass (environment) on the evolution of dwarfs (cf. Section\,\ref{sect:environment}). The hockey-puck plot of Fig.\,\ref{fig:spatial} (right) again compares the two samples, where the classic Fingers of God effect is clearly visible.

\subsection{Imaging and spectroscopic data}
One of the main goals of AVOCADO is to construct and analyse the UV-to-NIR spectral energy distribution (SED) of dwarf galaxies in the Local Universe. In spite of the wealth of optical spectroscopic information currently available thanks to recent large-area surveys (mainly the 2dFGRS and the SDSS), SEDs still provide invaluable and complementary information. It is well known (e.g., \citealt{Thuan1985,GildePaz2002}) that for nearby galaxies NIR photometry  can sensitively constrain the age, metallicity, and mass of stellar populations -- but the situation is more complex in the case of strongly star-forming galaxies \citep[see][]{Noeske2003}. Alternatively, UV data provide information on the amounts of dust, the age and mass, and the star formation rate of galaxies on a broader time scale than traced by H$\alpha$ emission. Furthermore, it has been shown that the UV emission provides reliable SFRs for galaxies with weak or non-existing H$\alpha$ emission, and that the latter increasingly underpredicts the FUV-derived SFRs for low-mass galaxies \citep{Lee2009a}. Additionally, for a fixed number of photometric points, wider wavelength baselines can significantly reduce the uncertainties in the derived parameters \citep{GildePaz2002}. Finally, the analysis of SEDs is not restricted to objects with SDSS spectra, and thus allows for a direct comparison with dwarfs at higher redshifts, where spectroscopic data are scarce, but multiwavelength imaging is abundant.

\begin{figure}
\centering
\includegraphics[width=.5\textwidth,clip=true]{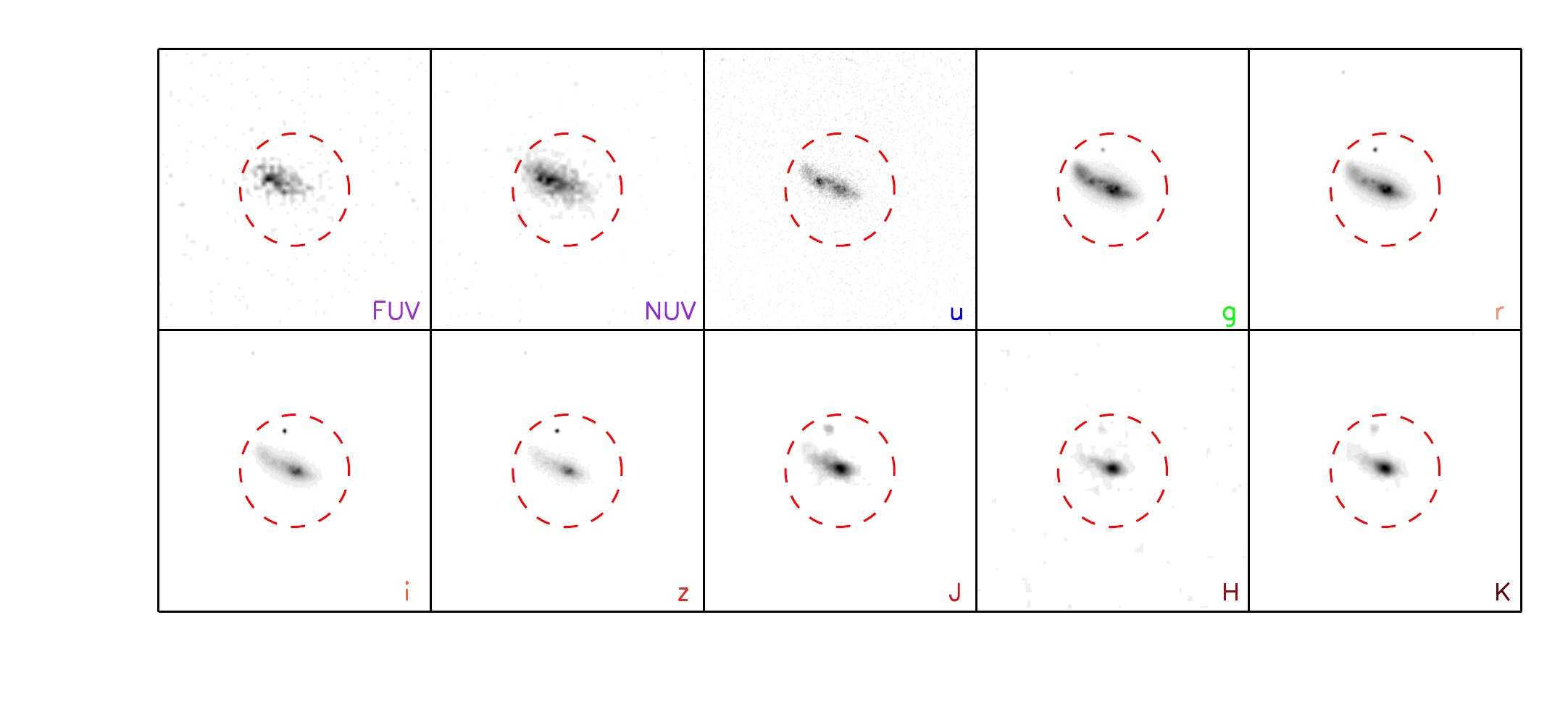} 
\caption{Example of postage-stamp images for a dwarf in the ten UV-to-NIR filters used in AVOCADO. The dashed circle indicates 2\,r$_{petro}$, the common aperture used to compute all fluxes. Note how the surface brightness distribution varies from the shortest to the longest wavelengths.} 
\label{fig:stamps}
\end{figure}

To this end we have retrieved GALEX+SDSS+2MASS  postage-stamp images centred on each object (see Fig.\,\ref{fig:stamps}) in our sample. 
Instead of using the photometric parameters derived by each of the surveys, we use these images to compute Petrosian magnitudes within a common aperture for all filters, which additionally allows measurements below the GALEX and 2MASS catalogue limits.

To retrieve this large amount of data we have taken advantage of the Virtual Observatory\,\footnote{\tt http://www.ivoa.net} (VO). The VO is a project designed to provide the astronomical community with the data access and the research tools necessary to enable the exploration of the digital, multi-wavelength universe resident in the astronomical data archives. In particular, we made use of Aladin\footnote{\tt http://aladin.u-strasbg.fr/} \citep{Bonnarel2000}, a VO-compliant software that allows users to visualise and analyse digitised astronomical images, and superimpose entries from astronomical catalogues or databases available from the VO services.

Thus, using Aladin in script mode, we proceeded with the following workflow for each galaxy. First, we downloaded an SDSS $u$-band image from the SkyView\footnote{\tt http://skyview.gsfc.nasa.gov} image server. This image is centred on the galaxy and has an angular size that varies from object to object. Specifically, each postage stamp has a half-side size of 4\,r$_{90}$, with r$_{90}$ computed as the average of the $g$, $r$, and $i$ radii. This size represents the best choice, because it provides an area large enough to accurately determine both the galaxy flux and the local sky background in the surroundings while minimising the number of (unwanted) stars and background galaxies present in the image.
Secondly, we loaded the SDSS-DR7 ($u$, $g$, $r$, $i$, and $z$-bands) images, the 2MASS ($J$, $H$, and $K_s$ bands) images, and the GALEX-GR4 ($FUV$ and $NUV$ bands) background-subtracted images of the same region of the sky. Since GALEX images are not available through the VO, we previously retrieved them from the GALEX server\footnote{\tt http://galex.stsci.edu} and locally loaded them with Aladin. And, finally, all images were resampled using the image from the SkyView as reference. 
%Consequently, we have ended up with a collection of GALEX+SDSS+2MASS postage-stamp images centred in the objects, which size has been set to 4\,r$_{petro}$ in order to produce a reliable estimation of both the object flux and its local sky background.
%{\bf I'm still fine-tuning the image treatment procedure and the photometric calibration of the data, so no 2MASS or GALEX photometry yet. In this section I will include three plots comparing our photometry in NUV, r and K with that of the GALEX, SDSS and 2MASS catalogues (for objects in common). }

In addition to the imaging data, we use the optical spectra provided by the SDSS. These consist of $R \approx 2000$ spectra in the $3800-9200$ \AA\ wavelength range, with a median $S/N=10$ for dwarfs in the AVOCADO sample. This value is similar to the main galaxy sample spectra analysed by \citet{Tremonti2004}.

%________________________________________________________________

\section{VOSA: the VO SED Analyzer}
\label{sect:VOSA}
Our ten-point UV-to-NIR SEDs are fitted to a library of single stellar populations (SSPs) using VOSA\,\footnote{\tt http://svo.cab.inta-csic.es/theory/vosa/}, a VO tool of public use.
The application was originally designed for the analysis of stellar objects \citep{Bayo2008} and a new independent workflow for extragalactic objects has been developed for this project -- taking into account the different physics that must be applied, using different VO photometric services and theoretical models, and implementing new capabilities (Bayo et al. 2012, submitted).
The tool includes two options to estimate the optimal model(s) reproducing the data: a traditional $\chi^{2}$ minimisation, and a Bayesian approach (in the spirit of \citealt{Kauffmann2003}). 

While any VO-compliant stellar population model can in principle be used in the fitting procedure, AVOCADO uses the PopStar evolutionary synthesis models of \citet{Molla2009}. 
The basic grid is composed by a variety of initial mass functions (IMFs), based on an updated version of the isochrones by \citet{Bressan1998} for six different metallicities: $\rm Z= 0.0001$, 0.0004, 0.004, 0.008, 0.02 and 0.05.  The use of very low metallicity models of $\rm Z=0.0001$ is a particularly relevant feature for the study of low-metallicity starburst dwarfs and, to our best knowledge, has not been included before in similar works. Age coverage ranges from $\log{t}=5.00$ to 10.18, with a variable time resolution of $\Delta(\log{t})=0.01$ in the youngest stellar ages.
For these youngest ages, the SEDs include the emission of H and He nebular continuum. This emission has a significant impact on ultraviolet, optical, and near-infrared colours during star formation activity -- especially for young, metal-poor stellar populations --, thus affecting their derived properties \citep{Zackrisson2001}.
We refer the reader to \citet{Molla2009} for more specific details concerning the evolutionary synthesis models.
The detailed analysis of UV-to-NIR SEDs and results from their fits for the AVOCADO sample will be presented elsewhere.

%Atmosphere models are from \cite{lcb97} with an excellent coverage in effective temperature, gravity and metallicities, for stars with Teff $\leq 25000$K.  For O, B and WR the code uses the NLTE blanketed models by \cite{snc02} at $\rm Z = 0.001$, 0.004, 0.008, 0.02 and 0.04. There are 110 models for O-B stars, calculated by \cite{phl01}, with 25000K $<$ Teff $\leq 51500$K and $2.95 \leq \log{g} \leq 4.00$, and 120 models for WR stars (60 WN and 60 WC), from \cite{hm98}, with 30000K $\leq T^{*} \leq 120000$K and $1.3R_{\odot}\leq R^{*}\leq 20.3 R_{\odot}$ for WN, and with $ 40000K \leq T^{*} \leq 140000$K and $ 0.8R_{\odot}\leq R^{*}\leq 9.3 R_{\odot}$ for WC. T$^{*}$ and R$^{*}$ are the temperature and the radius at a Roseland optical depth of 10. The assignation of the appropriate WR model is consistently made by using the relationships among opacity, mass loss and velocity wind as described in Paper I.  For post-AGB and PN with Teff higher than 50000K and up to 220000K the NLTE models by \cite{rau03} are taken. For higher temperatures {\sc PopStar} uses black bodies. The use of these last models modify the resulting intermediate age SEDs.

%________________________________________________________________

\section{Analysis of SDSS spectra}
\label{sect:spectra}
The analysis of SEDs will provide physical information for the whole sample of dwarfs, allowing for a detailed study of close companions and an optimal comparison with future, high-redshift studies that most probably will lack spectroscopic information. However, SDSS spectra are available for most of the AVOCADO sample, and we will take full advantage of this wealth of information.

\begin{figure}
\includegraphics[width=.5\textwidth]{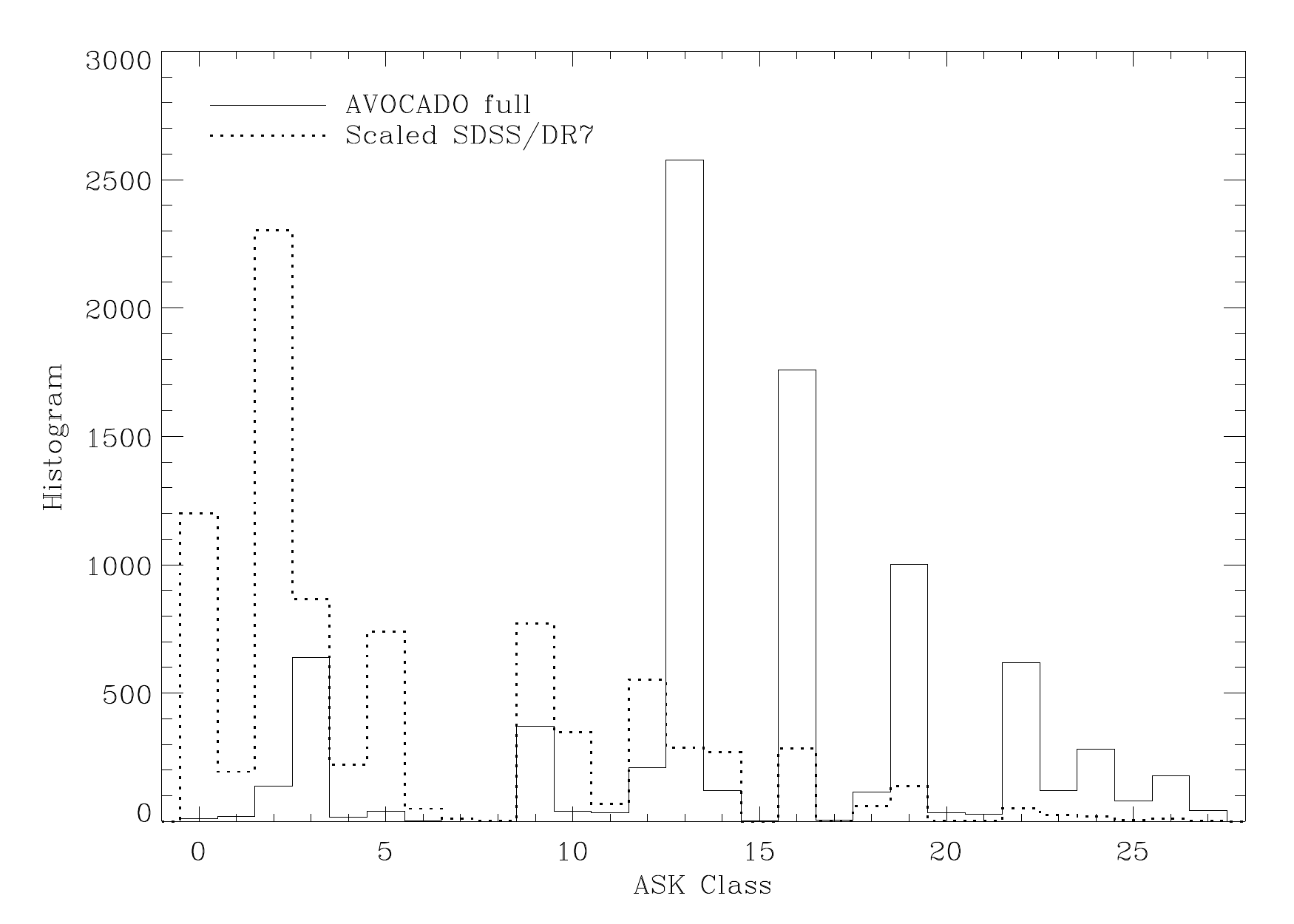}
\caption{
Distribution of spectroscopic classes that the AVOCADO dwarfs belong to (ASK classes; the
solid line). Most of them are star-forming galaxies (ASK$\,\geq\,7$), in sharp
contrast with the original SDSS/DR7 sample, whose distribution is shown as a
dotted line.
}
\label{histos}
\end{figure}

Galaxy spectra show a high degree of uniformity, and can therefore be assigned to a limited number of representative classes -- probably forming a
bifurcated one-dimensional sequence analogous to the morphological-type one \citep[e.g.,][]{Ascasibar2011}. 
\citet{SanchezAlmeida2010} carried out a clustering classification of all  galaxies with spectra in the SDSS/DR7. Their algorithm guarantees that galaxies with similar spectra belong to the same class, and they find that only 17 major classes describe $\sim$\,$10^{6}$ spectra (ASK classes hereafter; see Fig.\,\ref{histos}).\,\footnote{The ASK classification for all SDSS-DR7 galaxy spectra can be accessed at {\tt http://sdc.cab.inta-csic.es/ask/index.jsp}} They
broadly correspond to red, quiescent galaxies\footnote{Prototypes for elliptical galaxies,
but including a large fraction of red-spirals as well
\citep{SanchezAlmeida2011}.} (ASK\,0, 2 and 3), dust-shrouded edge-on spirals (ASK\,1
and 4), green-valley galaxies (ASK\,5), AGN-hosting galaxies (ASK\,6), as well as
a number of classes corresponding to star-forming galaxies (ASK\,7 and larger).

\begin{figure*}
\includegraphics[width=.9\textwidth]{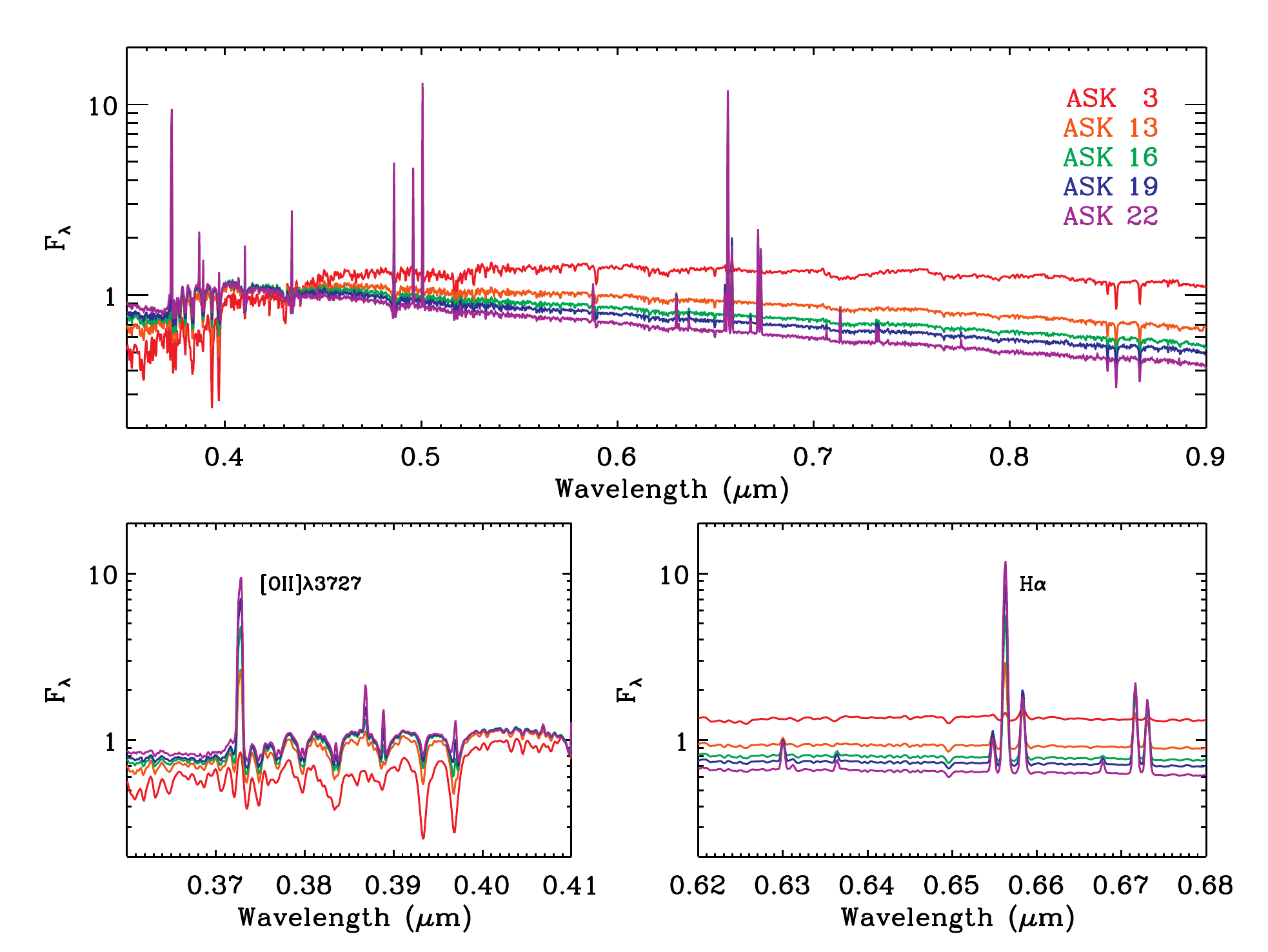}
\caption{
Average spectra of the five main spectral classes that AVOCADO galaxies belong to. Upper panel: the full optical spectral range. Lower panels: zoom into two particular spectral ranges, including {\sc [\oii]}$\lambda$\,3727 and H$\alpha$. Wavelengths are given in $\mu$m. Note how ASK\,3 represents a class of quiescent dwarf galaxies.
}
\label{templates}
\end{figure*}

We have identified the ASK spectroscopic classes of the AVOCADO dwarfs with SDSS 
spectra and, not surprisingly, most of them belong to this tail of star-forming galaxies. The actual
distribution of classes is represented in Fig.~\ref{histos} with the solid line. It contrasts with 
the (apparent magnitude-limited) parent ASK sample, which is dominated by quiescent classes 
(ASK$\,\leq\,5$; the dotted line in Fig.~\ref{histos}).
Most AVOCADO dwarfs belong to five main classes: ASK\,3 (7.5\%), ASK\,13 (30\%), ASK\,16 (21\%), ASK\,19 (12\%), and ASK\,22 (7.5\%). 
The template spectra for these classes are shown in Fig.~\ref{templates}. The star-forming templates
have H$\alpha$ in emission with equivalent widths ranging from EW\,$\approx$\,10\,\AA\
(ASK\,13) to EW\,$\approx$\,80\,\AA\ (ASK\,22). Using the star formation rates from the
H$\alpha$ flux as calibrated by Kennicutt (1998), and the
mass-to-light ratio as modelled by Bell \& de Jong (2001), the
sSFR goes from $\approx$\,40 Gyr$^{-1}$, for ASK\,13, to $\approx$\,2 Gyr$^{-1}$, for ASK\,22. The inverse sSFR
is the time scale required to produce the observed stellar mass at the current 
SFR, and therefore measures the strength of the starburst.
The majority of AVOCADO dwarfs are characterised by their low-level star formation activity,  with only a relatively small fraction qualifying in the category of actual starbursts.
Figure~\ref{histos} also shows the existence of AVOCADO dwarfs belonging to ASK\,3, which is a class of quiescent galaxies with spectra dominated by strong absorption features and lacking emission lines.  These systems can be identified with the early-type dwarfs that populate the red sequence in Fig.\,\ref{fig:cmd}, and allow us to investigate which mechanisms are responsible for the star formation activity shut-off (cf. Sect.\,\ref{sect:environment}).

This clustering analysis of spectra shows that dwarfs exhibit a great variety of current, and probably past, star formation activities. Recovering their detailed SFHs is one of our main goals.

%
%
%%%%%%%%

\subsection{Starlight spectral synthesis}
Spectral synthesis has proven to be one of the most powerful tools available for extragalactic astronomy to investigate the stellar populations in galaxies. Among the existing population synthesis codes, STARLIGHT \citep{CidFernandes2005} is one of the few that are publicly available and is entirely dedicated to the recovery of the star formation histories of galaxies. STARLIGHT uses an "inverse" approach, where the best linear combination of SSPs, i.e. coeval stellar populations with homogeneous chemical abundances, is found. In order to be more realistic, the code also includes reddening as an uniform dust screen and two kinematical parameters (line-of-sight velocity dispersion and systemic velocity) using a Gaussian kernel function to smooth the modelled spectrum. The fitting procedure searches for the minimum $\chi^2$ and is carried out in the parameter space through multiple independent Markov Chain Monte Carlo iterations plus a simulated annealing technique.

The derived spectral synthesis parameters, such as the total stellar mass, dust extinction, mean stellar age and metallicity, SFH, star formation rates, and others, will then be used to have a better understanding of dwarf galaxy formation \citep[e.g.,][]{Gomes2011} and will be compared with those derived from photometric SED fitting to quantitatively address  the limitations imposed by these studies.

\begin{figure*}
\centering
\includegraphics[width=.5\textwidth,angle=-90]{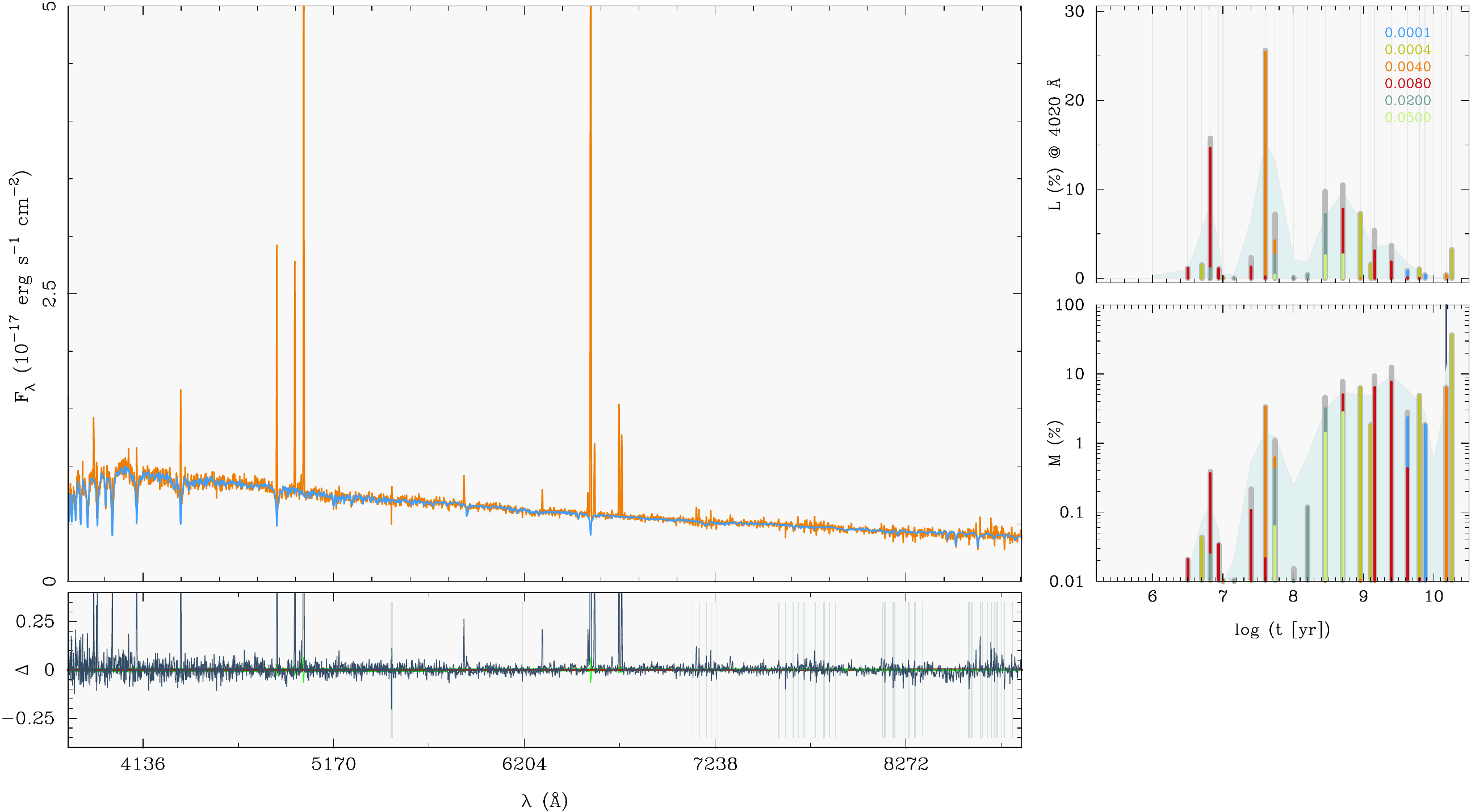} 
\caption{Example of STARLIGHT full-spectrum fitting for an AVOCADO star-forming dwarf galaxy. The top left panel shows the SDSS spectrum (orange), together with the best-fit model (blue) and its residuals (bottom left panel). The latter is a linear combination of SSPs featuring six metallicities and 25 age intervals. The corresponding SFH is shown in the rightmost panels, in terms of both light (top) and stellar mass (bottom) fractions. The colour-coding stands for the six different metallicities.} 
\label{fig:starlight}
\end{figure*}

In Fig.\,\ref{fig:starlight}, we show an example of spectral fitting with STARLIGHT for an SDSS star-forming dwarf galaxy, using a set of SSPs with six metallicities (0.005, 0.02, 0.2, 0.4, 1, and 2.5 Z$_\odot$) and 25 ages (between 1 Myr and 18 Gyr) from \citet{Bruzual2003}. This SSP library uses the Padova\,1994 evolutionary tracks \citep{Alongi1993,Bressan1993,Fagotto1994,Fagotto1994a,Fagotto1994b,Girardi1996} and a \citet{Chabrier2003} IMF between $0.1$ and $100\,\textrm{M}_\odot$. The reddening law was taken from \citet{Cardelli1989}. 
In the top left panel, the observed galaxy spectrum is presented (orange line), together with the best Starlight fit (blue line). The fitting residuals (observed -- modelled spectrum) are shown in the bottom left panel with previously masked regions, due to spurious pixels, in a light blue shaded area. In the right panels, we show the corresponding SFH for this galaxy in light fractions (top) at the normalisation wavelength (4020 \AA) and mass fractions (bottom). The colours stand for different metallicities (see label).
Clearly, even though the bulk of optical galaxy light is dominated by young ($<$100 Myr) stellar populations, the most significant contribution in terms of stellar mass comes from much older ($>$1 Gyr) stars (cf \citealt{Weisz2011}).

\subsection{The ionised gas}
Additionally, the continuum-subtracted spectra will allow the 
study of emission line equivalent widths (EWs) and line ratios 
to characterise the properties of the ionised gas in dwarfs, as 
well as to study the luminosity/mass-metallicity relation and 
other fundamental relations (e.g. between stellar mass, SFR, 
metallicity, and other ionic ratios such as nitrogen-to-oxygen) 
at the faint-end of the luminosity function. 
These relationships will allow us to additionally constrain  
the evolutionary status of the AVOCADO galaxies.

After subtracting the stellar continuum, emission line-integrated fluxes and EWs with their uncertainties, are obtained from the 
MPA/JHU Catalogue of the SDSS/DR\,7\footnote {Available at http://www.mpa-garching.mpg.de/SDSS/} \citep{Brinchmann2004,Tremonti2004}.
To properly measure physical quantities, we kept the 
non-duplicated galaxies with a signal-to-noise ratio (S/N) of at 
least 5 in all involved lines. 
%Emission-line integrated fluxes and EWs and their uncertainties are consistently measured for each galaxy spectrum using our own, %specifically designed, semi-automatic routines. 
%These consist  we fit Gaussians to the emission line after subtraction of the stellar continuum obtained by fitting the synthetic spectra by {\sc starlight}. 
Each emission line flux is corrected for the absorption of 
interstellar dust using the reddening constant $c(H\beta)$ obtained 
from the direct comparison between the observed \ha/\hb\ ratio and 
its theoretical value for standard physical conditions of the gas, and
assuming the extinction curve of \citet{Cardelli1989} as before.

To derive gas-phase metallicities for the whole sample of galaxies 
we use a combination of several strong-line methods involving the 
[\nii]\,$\lambda$\,6584\,\AA\ emission line.
This choice is motivated by the absence of the [\oii]\,$\lambda\lambda$\,3726, 3729\,\AA\AA\ emission lines in virtually all  galaxies (those with $z\la 0.02$, $\sim$99\% in our sample). 
This, together with the lack of trustworthy measures for the temperature-sensitive 
[\oiii]\,$\lambda$\,4363\,\AA\
 auroral line for a large part of the sample, 
prevents the use of the more reliable $T_{\rm e}$ method, in which 
the [\oii] line is required to calculate the contribution of the 
ion O$^+$ to the total oxygen abundance. 
We apply the so-called N2-method, which uses the strong-line 
calibrator N2 $\equiv$ \mbox{log([\nii]\,$\lambda$\,6584\,\AA/\ha)} [e.g., \citealt{Storchi-Bergmann1994,vanZee1998,Denicolo2002,Perez-Montero2005}], and has been extensively used 
to derive gas-phase metallicities in galaxies \citep[e.g.][]{Cairos2007,Lopez-Sanchez2010,Amorin2010,Morales-Luis2011}.
In particular, we use the empirical relation derived by \citet{Perez-Montero2009} 
between N2 and the oxygen gas-phase abundance. 
This relation gives metallicities fully
consistent with those derived from $T_{\rm e}$-sensitive methods over
most part of the range of expected metallicities (12+log(O/H)~$\ge$~8.0) 
[\citealt{Perez-Montero2009}, see also \citealt{Lopez-Sanchez2010}]. 
Furthermore, this relation does not present any dependence on reddening
correction or flux calibration uncertainties.
Additional empirical parameters involving [\nii], e.g., 
O3N2 ($\equiv$[\oiii]\,$\lambda\lambda$\,4959, 5007\,\AA\AA/[\nii]\,$\lambda$\,6584\,\AA; \citealt[][]{Alloin1979,Pilyugin2004}) and NS ($\equiv$[\nii]\,$\lambda$\,6584\,\AA/[\sii]\,$\lambda\lambda$\,6717, 6731\,\AA\AA; \citealt[][]{Kewley2002,Pilyugin2011}), will allow us 
to study the consistency of the results in different metallicity regimes. 
%Especially for the low metallicity regime (i.e. 12+log(O/H)$\lt$8.0), 
%where N2 may provide larger uncertainties, we will compare results 
%from N2 with those derived from strong-line methods involving the 
%[\oii] line, e.g., $R_{23}$ \citep{Pagel79,Pilyugin2001a,Pilyugin2001b,PMD03}. 
%In doing so we will use a very limited subset of galaxies ($\sim$50) located 
%in the high-end of sample redshift distribution, for which reliable 
%measurements of the [\oii] line are available. 

The nitrogen-to oxygen (N/O) ratio is another quantity of interest 
for AVOCADO because it has been shown that N/O is a powerful evolutionary 
indicator in galaxies \citep{Pilyugin2004,Koppen2005,Molla2006,Perez-Montero2009,Amorin2010,Amorin2012}.  
To derive the N/O ratio over the full range of metallicity values,  
strong-line calibrators depending on the ratios between [\nii] and 
other low-excitation lines -- like [\oii] and [\sii]  
(N2O2 and N2S2, respectively) -- can be safely used \citep{Perez-Montero2009}. 
For our study we decided to derive N/O using the N2S2 parameter and 
the last empirical calibration derived by \citet{Amorin2010} using a large 
sample of SDSS star-forming galaxies from the MPU/JHU catalogue with 
direct estimation of the ionic abundances. 
This choice also minimises possible dependences on reddening or flux 
calibration. 

%Following a similar procedure as for metallicities,  the additional use of other calibrations will provide a more realistic  view of the uncertainties and possible biases in the derivation of N/O.  Thus, we also use the N2O2 and N2S2 parameters with a different empirical calibration, previously obtained by \citep{Perez-Montero2009}. These calibrations are also based on galaxies in the whole range of masses with direct estimation of the ionic abundances. 

%________________________________________________________________

\section{Morphology and structure}
\label{sect:morph}

Ideally, one would like to compare the physical properties derived from spectra and SED fitting for the different types of known dwarfs. 
Dwarfs form a complex zoo of objects, but, as with more massive systems, they can be divided into two broad groups of early- and late-type systems. The former class consists mainly of normal and nucleated dwarf ellipticals (dEs) with a smooth morphology, and a probably increasing fraction of dwarf spheroidals (dSphs) in its faint luminosity and surface brightness end \citep[e.g.,][]{Binggeli1993,Mateo1998,Jerjen2012}. Because of selection effects, dSphs, as well as other intrinsically faint species of the early-type dwarf population (i.e. galaxies classified as ultra-faint and ultra-compact; e.g. \citealt{Jerjen2012}) are expected to be underrepresented in AVOCADO.

Late-type dwarfs, on the other hand, are generally characterised by their irregular, clumpy appearance, which is usually related to the presence of SF knots. Dwarf irregulars (dIs), dwarf spirals (dSp), and blue compact dwarfs (BCDs) are the most common examples of these systems. Surface photometry studies indicate that SF activity in BCDs and dIs is taking place within a more extended, underlying host galaxy \citep[e.g.][]{Loose1986,Papaderos1996,vanZee2000}. These two classes do not only differ in the amplitude of their ongoing SF activity, but also in the structural properties of their underlying host, with the host central density in BCDs exceeding  that in dIs by $\sim$1~dex \citep{Papaderos1996a}.

%Ideally, one would like to compare the physical properties derived from spectra and SED fitting for all the different types of known dwarfs.
%Dwarfs conform a complex zoo of objects, but, as with more massive systems, they can be divided into two broad groups of early- and late-type systems. The former comprise dwarf ellipticals (dEs) and spheroidals (dSph), featureless galaxies with smooth morphologies. The latter, on the other hand, are characterised by their irregular, clumpy appearance, usually related to the presence of SF knots. Dwarf irregulars (dIs), dwarf spirals (dSp) and blue compact dwarfs (BCDs) are the most common examples of these systems.

The assignment of dwarfs into any of these  types has been traditionally carried out through visual inspection of optical images, but this is a slow and subjective procedure. If galaxies can be visually classified into different morphological types, it is only because they occupy (more or less) delimited regions in a given parameter space of observables (e.g., \citealt{GildePaz2003}). 
In this spirit, every galaxy in AVOCADO is characterised by a set of quantitative morphological parameters -- including asymmetry, clumpiness, concentration, M$_{20}$, Gini, $b/a$ and $<$$\mu$$>_{e}$, among others. We note that colour is purposely \emph{not} used as a discriminator to decouple the galaxy structural information from its current star formation activity.\,\footnote{Of course, structural parameters also depend on the luminosity output from the star-forming component. They are therefore also affected by SF activity, and in a different amount for, e.g., weakly star-forming galaxies and BCDs. However, with our approach we derive the relation between SF activity and structure \emph{a posteriori}, by not imposing any prior dependence on colour.}

\begin{figure*}
\label{fig:morph}
\begin{center}
\begin{tabular}{ccccc}
\includegraphics[scale=.45]{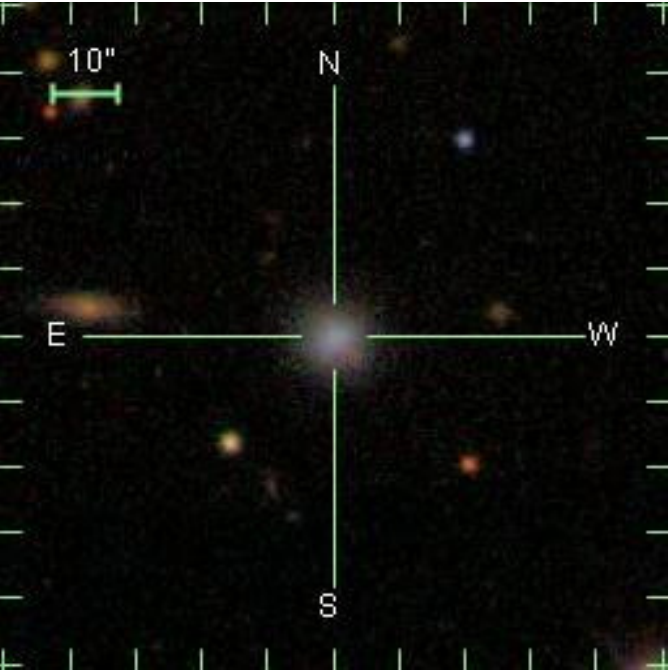} & \includegraphics[scale=.45]{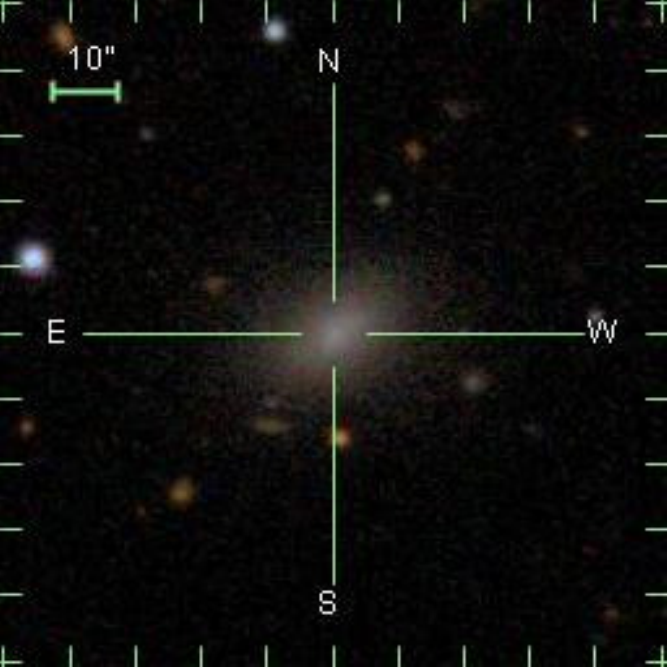} & \includegraphics[scale=.45]{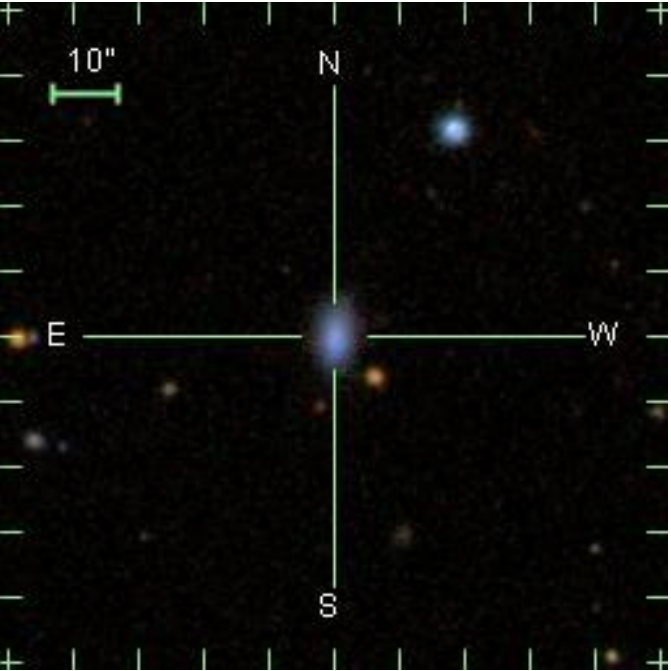} & \includegraphics[scale=.45]{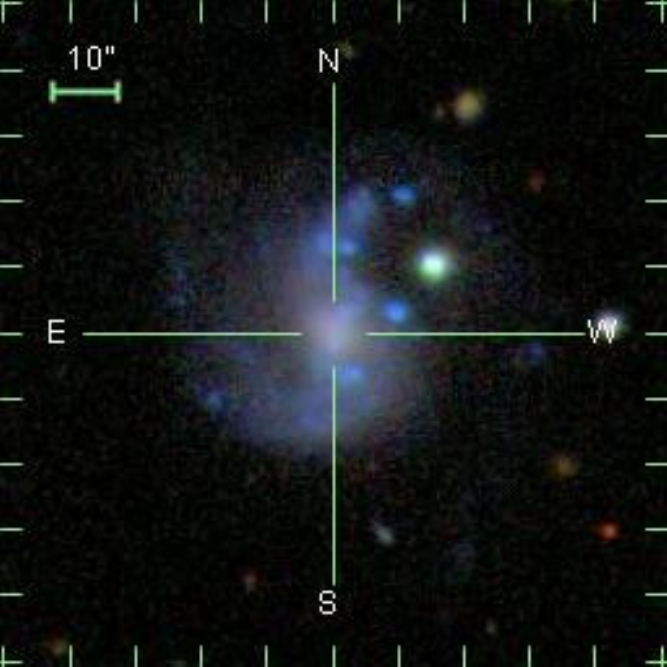} & \includegraphics[scale=.45]{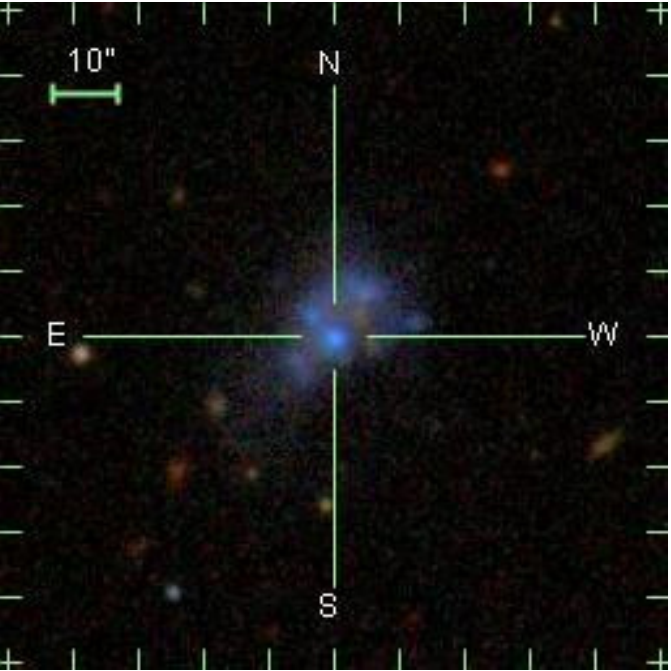}\\
%late diffuse irregular & late compact  \\
\includegraphics[scale=.45]{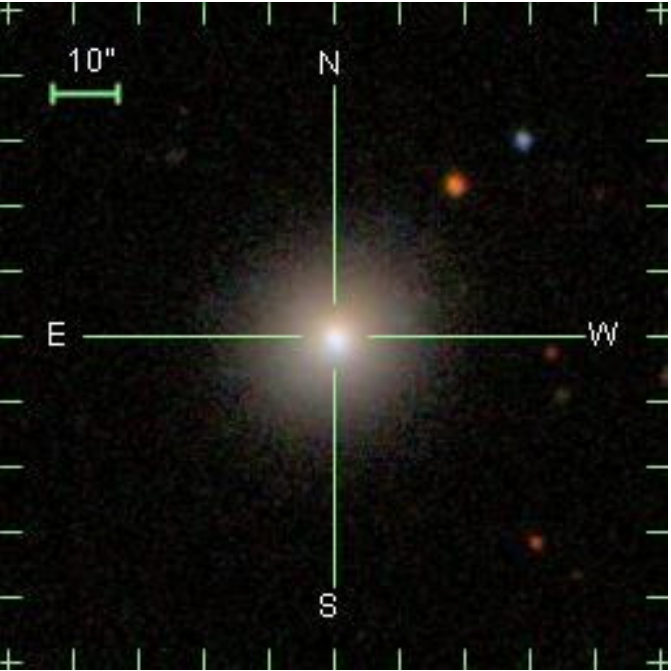} & \includegraphics[scale=.45]{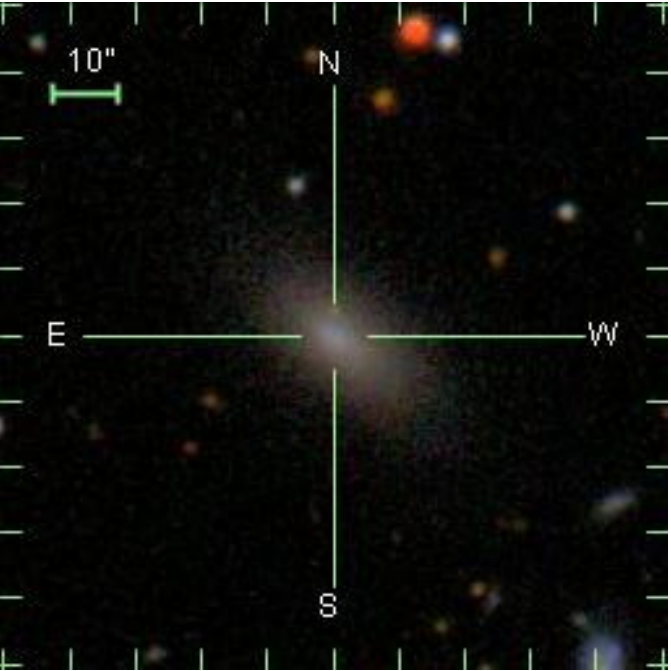} & \includegraphics[scale=.45]{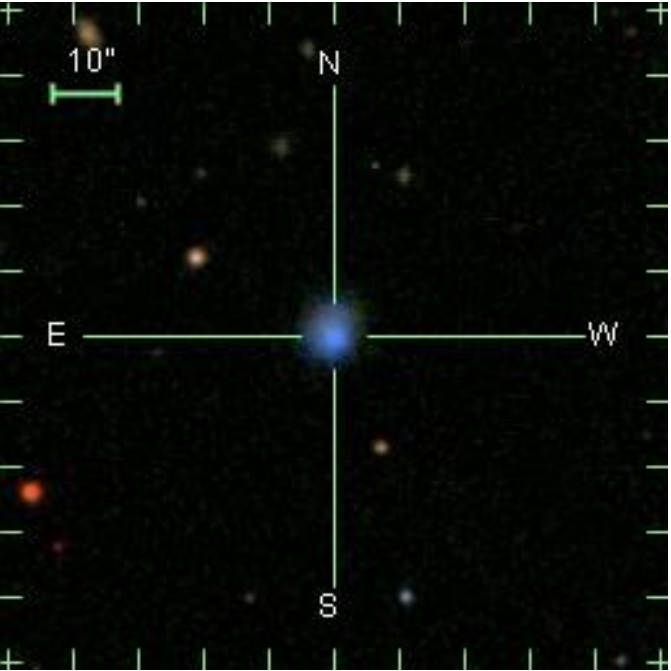} & \includegraphics[scale=.45]{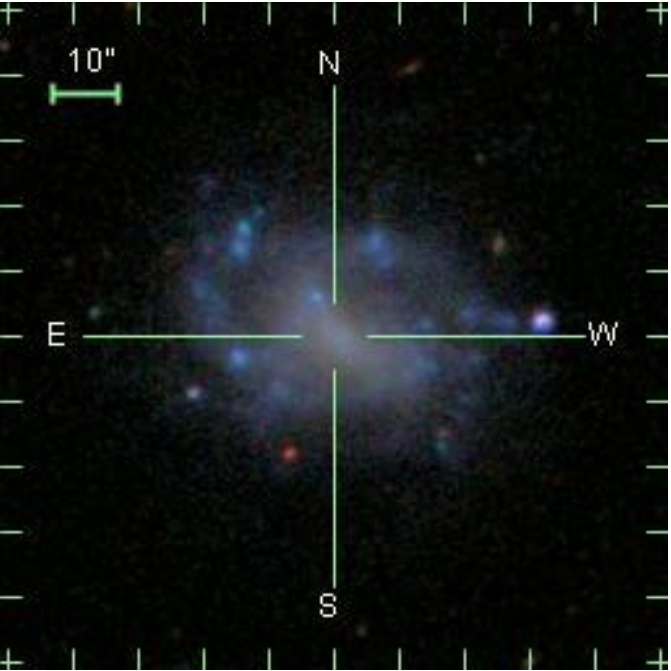} & \includegraphics[scale=.45]{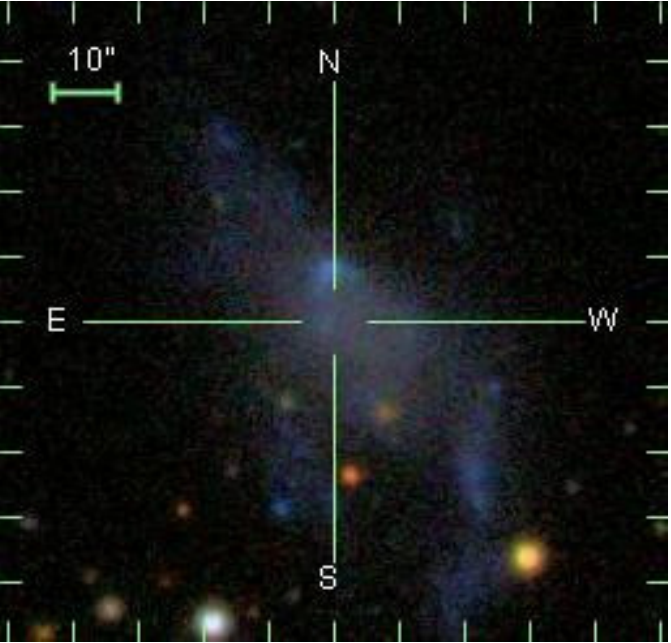}\\
Compact Early-Type  & Diffuse Early-Type & Compact Late-Type & Diffuse Spiral Late-Type &  Diffuse Irregular Late-Type\\
\end{tabular}
\end{center}
\caption{SDSS $gri$ postage-stamp images with two examples of each morphological type used in the AVOCADO training set. The assigned (visual) morphological classification is indicated in each case. All systems are located at a distance $\sim$\,20 Mpc, and therefore each image is roughly $\sim$\,$10\,h_{100}^{-1}$ kpc a side.}
\end{figure*}

To facilitate the comparison between different morphological types and with previous results, we again follow a statistical approach. The galSVM code \citep{Huertas-Company2008} uses support vector machines (SVM) to identify non-linear boundaries in the previous $n$-dimensional parameter space. Boundaries are defined using a training set that is built from a subsample comprising the closest 1500 dwarf galaxies in the AVOCADO sample. 
The power of this approach  is that by using a data subsample in the SVM training, the whole classification is fully data-driven. 
This training set was visually classified independently by three of us (RSJ, RA, MHC). For this purpose, for every galaxy we first determined whether its attributes are characteristic of early (E) or late-types (L). Within the first group, we furthermore distinguished between compact (C) and diffuse (D) systems. Late-types were also classified according to this criterium, but another level was introduced for the latter: diffuse dwarfs were subdivided into irregular (I) and spiral-like (S) objects. 
We stress that even though the visual classification was carried out on the SDSS $gri$ colour images, only morphological features were used to assign a galaxy to a given class. While some subjective colour information could be encoded in these visual classifications, it will not have any impact on the automated ones -- with our approach, galSVM cannot distinguish between two objects with different colour if they are structurally identical.
Figure\,\ref{fig:morph} shows some examples of galaxies in the training set, together with their corresponding morphological visual classification.

\begin{figure}
\centering
\includegraphics[width=.4\textwidth]{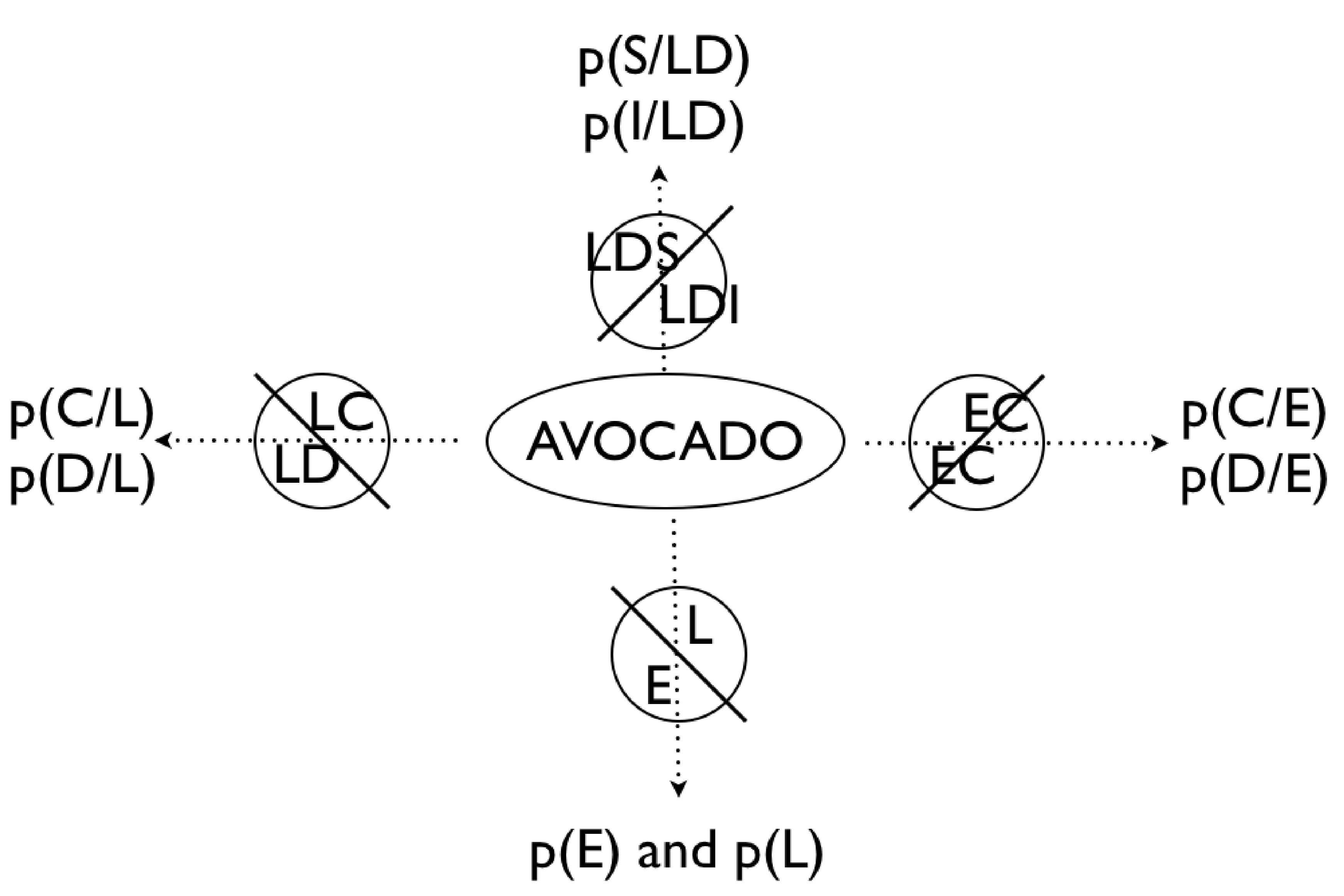} 
\caption{Schematic representation of the procedure used for the Bayesian morphological classification of galaxies in the AVOCADO sample (see text for details). The different labels correspond to early-types (E), late-types (L), compact (C), diffuse (D), spiral-like (S), and irregular (I) galaxies.} 
\label{fig:morpho_proc}
\end{figure}

We then ran galSVM and associated  six probabilities to every galaxy in our sample, according to its location in the parameter space:
 \begin{itemize}
 \item $p(C_E)$: probability to be compact early-type (dE-like).
 \item $p(D_E)$: probability to be diffuse early-type (dSph-like).
 \item $p(C_L)$: probability to be compact late-type (BCD/H{\sc ii}-like).
 \item $p(D_L)$: probability to be diffuse late-type.
 \item $p(S)$: probability to be dwarf spiral-like (dSp-like).
 \item $p(I)$: probability to be irregular-like (dIrr-like).
 \end{itemize}
 
Thus, the following relation holds for all AVOCADO dwarfs:  $p(C_E)+p(D_E)+p(C_L)+p(D_L)+p(S)+p(I)=1$.
Since SVMs work better in two-class problems, each of these probabilities was computed after several runs of galSVM with different training sets following a Bayesian approach (see Fig.\,\ref{fig:morpho_proc}), similar to what is explained in \citet[][]{Huertas-Company2011}.  Briefly, to obtain the probability $P(C_E)$ we first classified all galaxies into two classes (late and early), and then we created a training subset comprising early-types only -- which were additionally separated into the diffuse and compact classes. This second step returns the conditional probabilities $p(C|E)$ and $p(D|E)$. Finally, applying the Bayes theorem, $p(C_E)=p(E) \times p(C|E)$. Obviously, we also have $p(E)=p(D_E)+p(C_E)$. 

\begin{figure}
\includegraphics[width=0.5\textwidth]{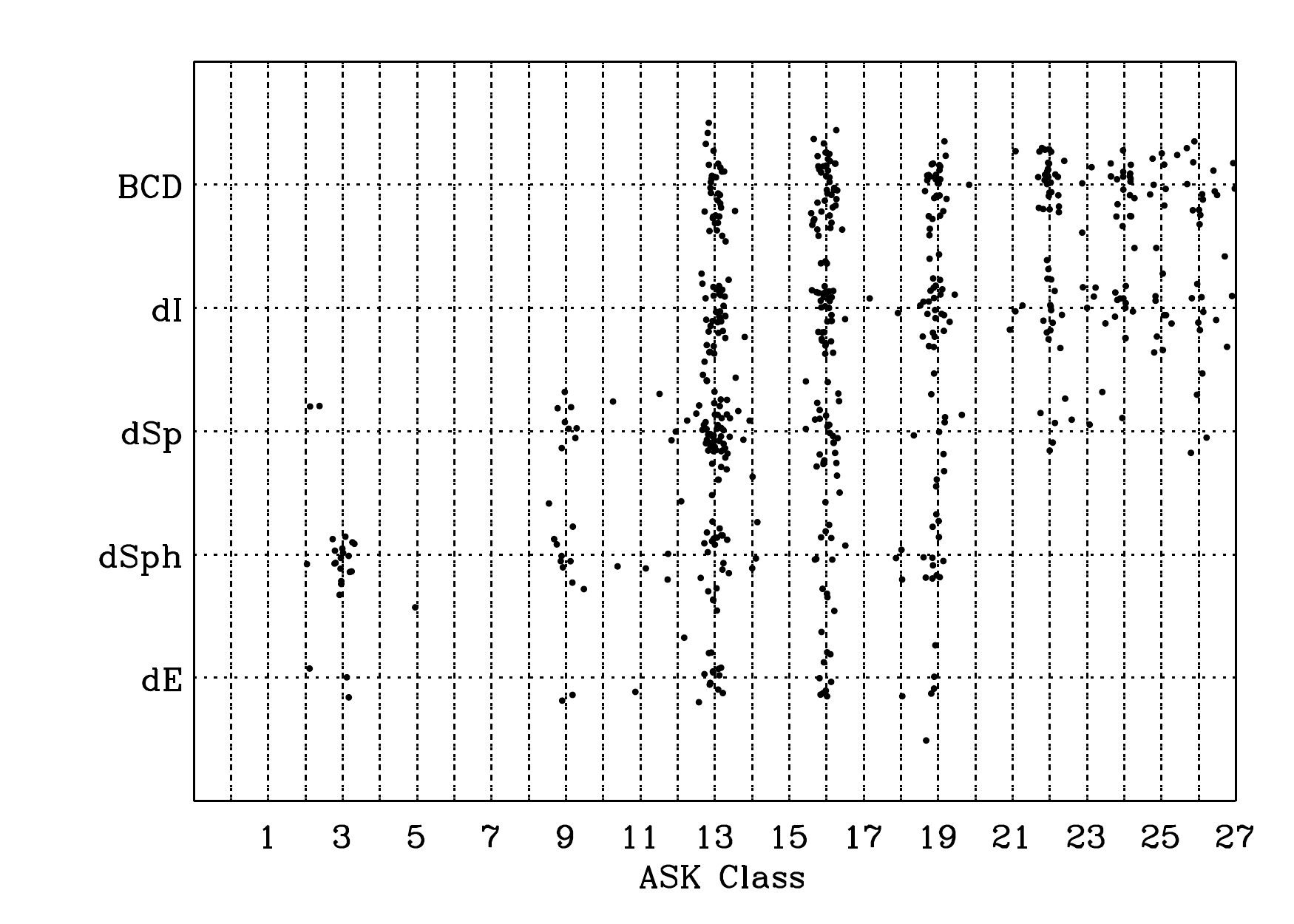}
\caption{Visual morphological type vs spectroscopic ASK class for the subsample of 1500 closest AVOCADO dwarfs used in the training sample. Note the expected trend that later-type galaxies are preferentially characterised by spectra with stronger starburst features, while earlier types are quiescent, or have moderate star formation activity.}
\label{hubble_vs_ask}
\end{figure}

Figure\,\ref{hubble_vs_ask} shows the relation between the (visual) morphological types of the 1500 galaxies in our training set, and the ASK class they belong to. It is obvious that there is no simple relation between both classifications, but the expected trends are nonetheless recovered. Later-type galaxies (BCDs and dIs) are preferentially characterised by spectra with stronger starburst features (ASK $>$ 7). Galaxies with more moderate star formation activity, however, can be found among almost all morphological types. Quiescent spectra (ASK $<$ 7) are only found in early-type systems.

In addition to this approach, and complementary to it, we will also derive and investigate the stellar surface brightness profiles of AVOCADO dwarfs, decomposing them into the luminosity contributions from the host and the star-forming component \citep[e.g.,][]{Papaderos1996a,Cairos2001a,Bergvall2002,Caon2005,Amorin2007,Amorin2009}. We will therefore obtain a complete and comprehensive view of the morphological and structural properties of dwarf galaxies in the nearby Universe.

%________________________________________________________________

\section{Environment}
\label{sect:environment}

\begin{figure*}
\centering
\includegraphics[width=.9\textwidth]{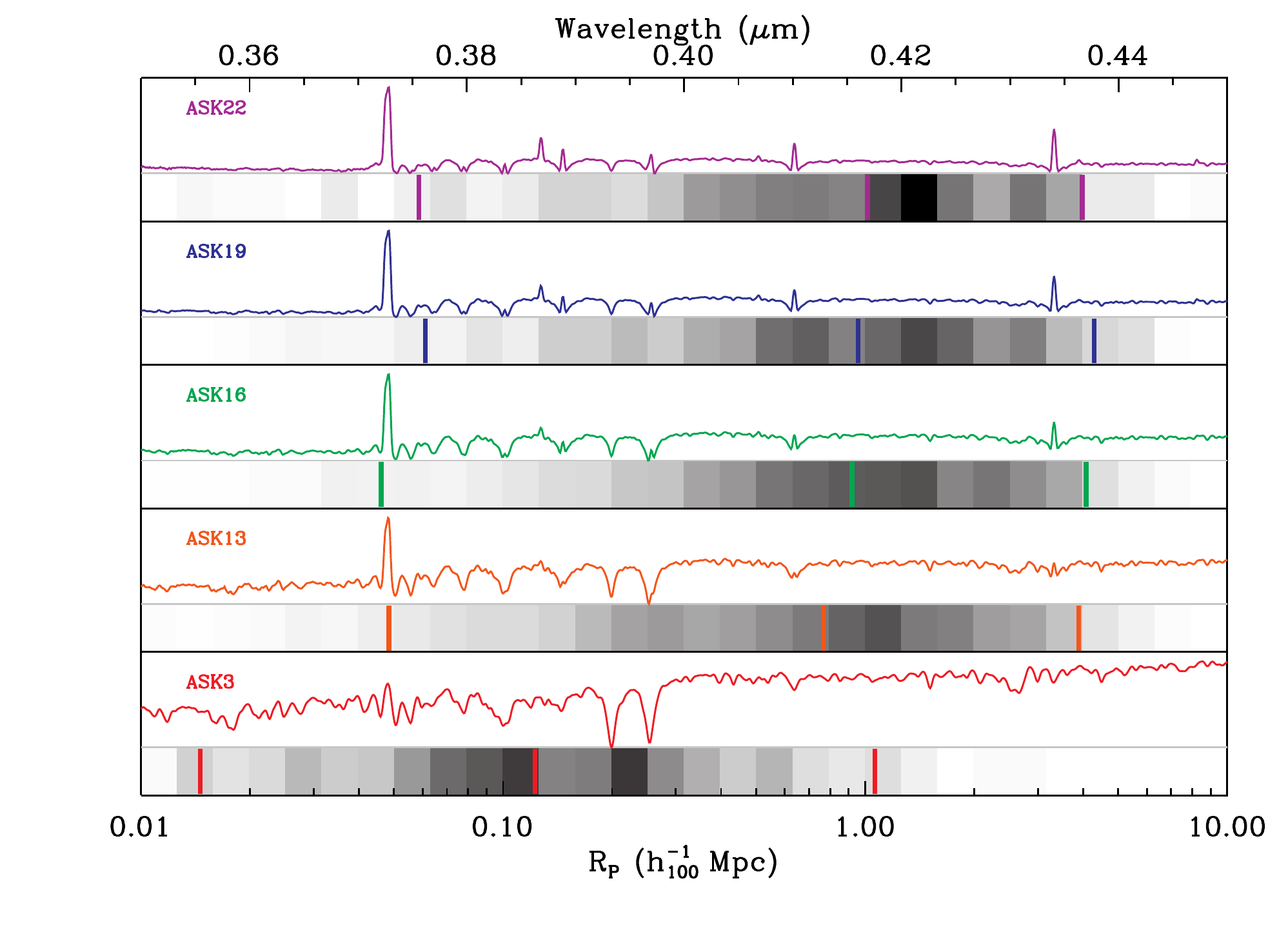} 
\caption{Grey scale shows, for AVOCADO dwarfs of each ASK class, the normalised distribution of (projected) distances to the closest luminous ($M_{K} < M_{K}^{*}+1$) galaxy in the 2MRS catalogue. The three vertical lines indicate, in each case, the 2.5, 50 and 97.5 per cent quantiles of the R$_{\mbox{\tiny P}}$ distributions, and the corresponding ASK template spectra are shown, in logarithmic flux scale, for reference. ASK classes are sorted, from top to bottom, in order of decreasing current star formation activity. We find a strong environmental trend for the non-star-forming ASK\,3 galaxies, which are only found closer than $\approx$\,1.5\,$h_{100}^{-1}$ Mpc from a luminous companion -- i.e., isolated quiescent dwarfs are essentially absent in the $20 < D < 60\,h_{100}^{-1}$ Mpc volume we probe.
%In other words, being a satellite of an $\sim$\,L$^{*}$ central is a necessary condition to create a quiescent dwarf galaxy.
} 
\label{fig:neighbour}
\end{figure*}

\begin{figure}
%\centering
\includegraphics[width=.5\textwidth]{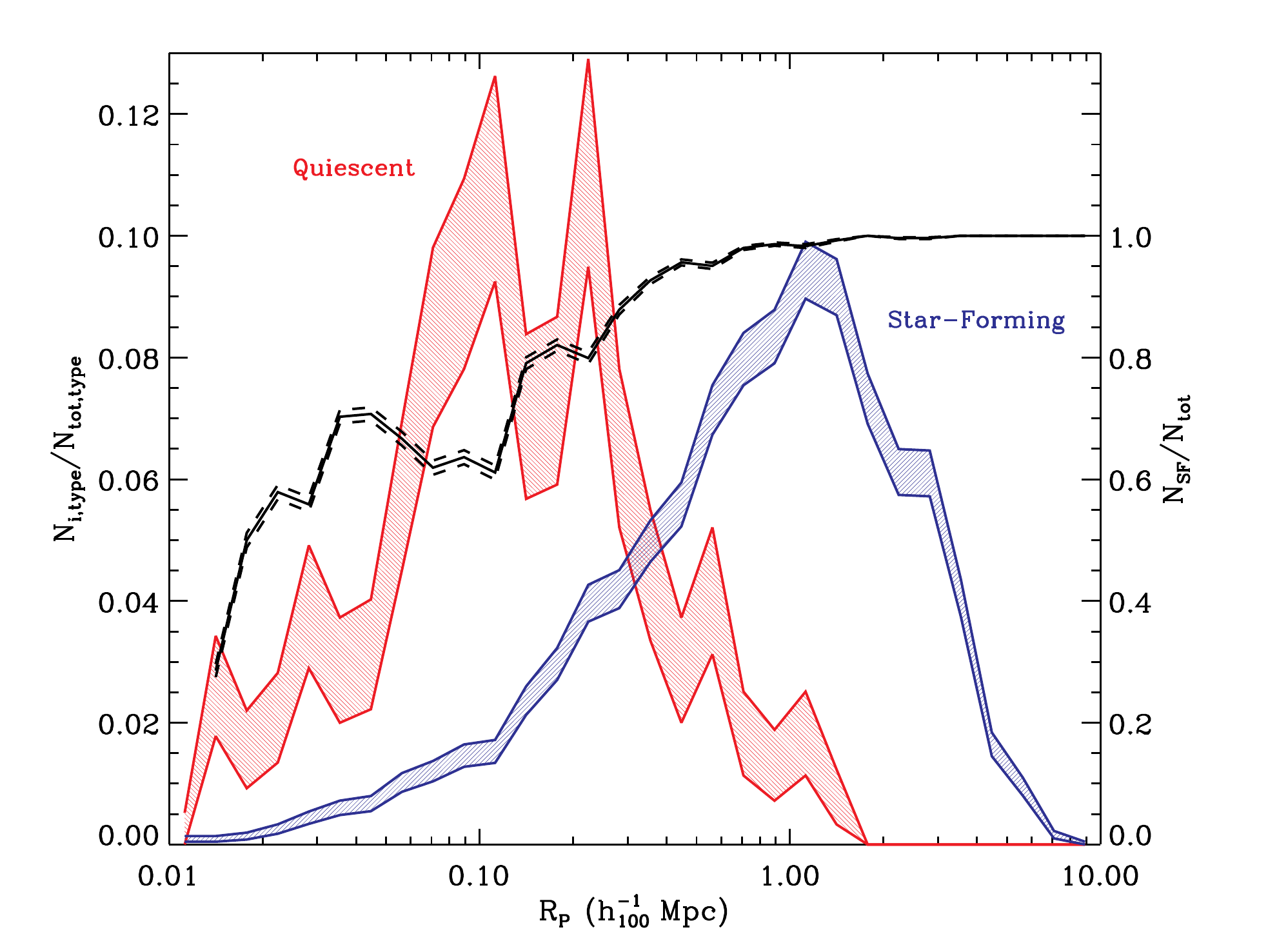} 
\caption{
Filled histograms show the normalised distributions of projected distances to the closest luminous galaxy in the 2MRS catalogue for quiescent and all star-forming dwarfs. The width of each histogram indicates the corresponding poisson uncertainty. It is clear that star-forming dwarfs are preferentially found at distances $\sim$$\,1\,h_{100}^{-1}$ Mpc from their hosts, while quiescent dwarfs almost never reach such large separations and generally concentrate in the inner $\sim$$\,100-200 \,h_{100}^{-1}$ kpc. The solid curve shows the fraction of star-forming galaxies at each galactocentric distance, with dashed lines indicating the 95\% binomial uncertainty interval. Star-forming dwarfs absolutely dominate for distances $\gtrsim$$\,400 \,h_{100}^{-1}$ kpc. Their abundance starts to noticeably decline at smaller projected distances, but they remain dominant down to separations as small as R$_{\mbox{\tiny P}} \sim 20\,h_{100}^{-1}$ kpc -- where quiescent dwarfs take over.
} 
\label{fig:neighbour2}
\end{figure}

The role that stellar mass and halo mass (environment) play in galaxy evolution is not yet fully understood. What is clear though is that central and satellite galaxies exhibit strikingly different properties, suggesting that environmental effects are indeed relevant. These might be even more important for dwarf galaxies, given their low masses and densities. Indeed, early-type dwarfs in the Local Universe tend to inhabit higher density environments than later types \citep[e.g.,][]{Binggeli1991,Vilchez1995,Karachentsev2004,Weisz2011}.

Recent studies (e.g., \citealt{Blanton2006})  suggest that  the effects of environment are predominantly local, i.e., the host halo mass is the main parameter. The satellite position within the host galaxy halo, or the presence of faint close companions, only introduce second-order effects -- at least in the case of relatively luminous galaxies. It is also well-established that among the different methods to estimate environment, the nearest-neighbour approach provides the most accurate description of the local density -- if the number of neighbours is sufficiently small to actually probe the halo \citep[e.g.,][]{Muldrew2012}. 
Unfortunately, it is well known that environmental studies with the SDSS suffer from severe limitations for nearby galaxies, such as edge effects, bright galaxy shredding and spectroscopic bias against close companions. 
We therefore adopt the following approach to characterise the environment of the dwarf galaxies in our sample.  
We use the 2MRS \citep{Huchra2011} catalogue to construct an all-sky sample of galaxies essentially complete down to $M_{K}-5\,\mbox{log}\,h_{100}  = -22.5$ mag within the volume probed by AVOCADO. This corresponds to $\approx$\,$M_{K}^{*}+1$ \citep{Kochanek2001,Bell2003}, guaranteeing that for all dwarfs, we include all companions with a mass ratio $\gtrsim$\,4:1 -- the classical limit for minor mergers. 
For each AVOCADO dwarf we use this luminous sample to look for the closest companion within $\Delta$v $\le 1000$ km\,s$^{-1}$ \citep[e.g.,][]{Blanton2006}.
Considering the intrinsic low luminosity of our objects and the lack of massive clusters in our sample, this approach ensures that we mainly probe the halo of the central galaxy our AVOCADO dwarf is a satellite of -- if any.\,\footnote{Shortly after this paper was first submitted, \citet{Geha2012} presented a very similar analysis, reaching the very same conclusions we detail below.}

Figure\,\ref{fig:neighbour} (grey scale) presents  the normalised distributions of projected distances, R$_{\mbox{\tiny P}}$, to the closest luminous galaxy in the 2MRS catalogue  for dwarfs in each of our five main ASK classes. The corresponding ASK template spectra are shown for reference, and the classes are sorted, from top to bottom, in order of decreasing current star formation activity. The three vertical lines indicate, for each ASK class, the 2.5, 50, and 97.5 per cent quantiles of the R$_{\mbox{\tiny P}}$ distributions.
There are several features worth discussing in Fig.\,\ref{fig:neighbour}.
First, it shows that star-forming dwarfs (ASK\,22, 19, 16, and 13) are preferentially found  at moderately large distances from luminous neighbours, with a median R$_{\mbox{\tiny P}} \sim 1\,h_{100}^{-1}$ Mpc -- but the projected distance range is significant, from several Mpc down to $\sim$$\,50\,h_{100}^{-1}$ kpc. 
Remarkably, the latter is roughly the distance between the Milky Way and the Magellanic Clouds, lending support to the argument that their high luminosity and blue colours --together with their proximity to their central galaxy-- make them rare systems among the satellite population of $\sim$\,$L^{*}$ galaxies \citep[see][]{James2011,Liu2011,Tollerud2011}.
Second, the median projected distance exhibits a mild trend with current star formation activity, in the sense that it decreases from R$_{\mbox{\tiny P}} \approx 1\,h_{100}^{-1}$ Mpc for dwarfs with the highest SFRs (ASK\,22), to R$_{\mbox{\tiny P}} \approx 750\,h_{100}^{-1}$ kpc for dwarfs with moderate SFRs (ASK\,13).
The strongest environmental trend, however, is found for the class of quiescent dwarfs: virtually none (0.75\%) of the ASK\,3 galaxies are found farther than $1.5\,h_{100}^{-1}$ Mpc from a luminous companion -- i.e., isolated quiescent dwarfs are essentially absent from the $20 < D < 60\,h_{100}^{-1}$ Mpc volume we probe.\,\footnote{One can of course argue that our spectral classification only refers to the galaxy central regions -- due to the limited spatial coverage of the SDSS fibres --  and that off-centre SF activity might still be going on in these quiescent dwarfs. This possibility is strongly ruled out by visual inspection of the $\sim$\,400 ASK\,3 dwarfs in our sample, but will be additionally quantified through the use of UV imaging and colour maps \citep[e.g.,][]{Cairos2001}.}

The same effect is also shown in Fig.\,\ref{fig:neighbour2}, where the filled histograms show, now for the two populations of quiescent (red) and all star-forming dwarfs (blue), the normalised distributions of projected distances to the closest luminous galaxy in the 2MRS catalogue. It is clear that star-forming dwarfs are preferentially found at distances $\sim$$\,1\,h_{100}^{-1}$ Mpc from their hosts, while very few quiescent dwarfs reach such large separations and mostly concentrate at the inner $\sim$$\,100-200 \,h_{100}^{-1}$ kpc. The solid line in the same figure shows the total fraction of dwarfs that are actively forming stars at each galactocentric distance, with dashed lines indicating the 95\% binomial uncertainty interval. Star-forming dwarfs absolutely dominate for distances $\gtrsim$$\,400 \,h_{100}^{-1}$ kpc. Their abundance starts to noticeably decline at smaller projected distances, but they remain dominant down to separations as small as R$_{\mbox{\tiny P}} \sim 20\,h_{100}^{-1}$ kpc -- where quiescent dwarfs take over.
We note that there appears to be a tail towards low separations in the distribution of R$_{\mbox{\tiny P}}$ for star-forming dwarfs, resulting in a flattening of their relative fraction for distances R$_{\mbox{\tiny P}} \lesssim 100\,h_{100}^{-1}$ kpc. We speculate that, modulo projection effects, this probably represents a genuine population of late-type dwarfs seen upon first infall about their host and before environmental quenching has fully operated -- pretty much as has been suggested to be the case for the Magellanic Clouds \citep{Besla2007}. 

All this suggests that being a satellite in the vicinity of a massive central is a \emph{necessary} condition to create a quiescent dwarf galaxy, in agreement with recent results \citep{Haines2007,Haines2008,Weisz2011,Geha2012}. 
Internal mechanisms --such as gas consumption via star formation or feedback effects-- appear to be insufficient to completely cease the star formation activity in dwarf galaxies. External processes need therefore to be invoked, and they most likely involve a combination of tidal and gas-stripping mechanisms \citep[e.g.,][]{Mayer2001,Mayer2001a,Grebel2003,Mayer2006,Penarrubia2008,Kazantzidis2011}.
One might argue that the fact that quiescent low-mass galaxies are found out to projected distances larger than 1 Mpc --corresponding to several virial radii for haloes hosting $L \gtrsim L^{*}$ galaxies-- is not consistent with the importance of environmental effects.
However, numerical simulations systematically show that there exists a non-negligible fraction of subhaloes located at large distances from the main halo that at some point of their history were inside its virial radius. This so-called backsplash galaxy population is known to exist in galaxy clusters \citep[e.g.,][]{Gill2005,Aguerri2010}, and several Milky Way satellites have been recently suggested to belong to this category \citep{Teyssier2012}. The numerical simulations by these authors show that more than 10\% of 'field' subhaloes have orbited through the virial volume of the Galaxy, and that they can be found at separations as large as 1.5 Mpc (or $\approx$\,5\,$R_{vir}$).
Therefore, the existence of environmentally quenched dwarfs at such distances is fully consistent with the expectations for the orbital evolutionary histories of satellites.

%________________________________________________________________

\section{Neutral and molecular gas content}
\label{sect:gas}

%\begin{figure}
%\centering
%\includegraphics[width=.5\textwidth]{avocado_HI.pdf} 
%\caption{Dots indicate all galaxies in the SDSS-DR7 (including luminous objects) that have an H\,{\sc i} detection in the Cornell H\,{\sc i} Data Archive. Notice the lack of detections at low masses, and how these naturally are concentrated within the blue cloud -- as opposed to the more luminous counterparts, where the red sequence also contains a significant number of detections. } 
%\label{fig:hi}
%\end{figure}

The final ingredient still missing to draw a robust evolutionary scenario for dwarf galaxies is their neutral and molecular gas content.
For AVOCADO we are compiling H\,{\sc i} data from the literature for a several hundred dwarfs, most of them arising from the ALFALFA survey \citep{Giovanelli2005}. %A cross-correlation with the Cornell H\,{\sc i} Data Archive reveals the only 242 dwarf galaxies of the AVOCADO sample that are detected in H\,{\sc i}. In Fig.\,\ref{fig:hi} we show the location of these galaxies in the CMD presented in Section\,2. As expected, most of the galaxies are located within the blue cloud, with very few objects on the red sequence -- but the opposite is not true for more luminous galaxies. Given that dedicated, deep observations have shown that dwarf galaxies contain most of their baryonic mass in the form of neutral gas, the lack of detections for faint objects has to be simply understood as an observational effect. 
\citet{Huang2012} show that gas-rich dwarfs are characterised by H\,{\sc i} discs that extend farther away than their stellar body and have relatively high sSFRs -- though the scatter is large at low masses. It is highly desirable to extend this kind of analysis to a well-characterised sample the size of AVOCADO.
A major breakthrough in the statistical studies of H\,{\sc i} properties of dwarf galaxies, however, will have to wait for the completion of already planned all-sky surveys, such as WALLABY and WNSHS. This is one of the long-term goals of our project.

Even though the great majority of field dwarf galaxies are actively forming stars, molecular gas -- the seed of this activity -- remains an elusive component \citep{Leroy2005}. It is still not clearly understood whether the lack of CO detection arises due to a true paucity of molecular hydrogen (H$_{2}$), or if CO becomes a poor tracer of H$_{2}$ at low metallicity environments.
We have started a programme aimed at observing a small subsample of AVOCADO dwarfs across a range of stellar masses and metallicities using the IRAM 30\,m antenna to determine their molecular content, in an effort to separate the joint contributions of these two fundamental parameters. Furthermore, large CO surveys for the bulk of the sample will be possible  in the near future with the recent advent of ALMA, with profound implications specially for these faint, metal-poor dwarfs.

%________________________________________________________________

\section{Summary}
We have presented a brief introduction to the AVOCADO project, including its science goals, the rationale behind the sample and dataset selection, and an overview of the main analysis tools. 
The main strength of AVOCADO is its unrivalled statistical power, built upon the homogeneity of the dataset and the uniform analysis methodology.
The sample consists of approximately 6500  nearby dwarfs ($M_{i}-5\,\mbox{log}\,h_{100} > -18$ mag), selected to lie within the $20 < \mbox{D} < 60~h_{100}^{-1}$ Mpc volume covered by the SDSS-DR7 footprint. It is thus volume-limited for $M_{i}-5\,\mbox{log}\,h_{100}<-16$ mag dwarfs, but includes $\approx$\,1500 fainter systems down to $M_{i}-5\,\mbox{log}\,h_{100} \approx -14$ mag. For all these systems we have compiled optical spectra and UV-to-NIR imaging, which are furthermore supplemented with structural parameters that are used to classify them morphologically, and a detailed characterisation of each dwarf's environment.

AVOCADO will provide a much-needed complement to the detailed studies carried out on dwarfs in the Local Volume, extending them to probe a greater variety of environments and including the most extreme dwarf types, which are poorly represented or simply missing in most Local samples.
In forthcoming papers of this series we will address the universal distribution of dwarf morphological types  (Huertas-Company et al.), the structural properties of their stellar component (Papaderos et al.), the properties of the ionised gas in star-forming dwarf galaxies (Amorin et al.), and the characteristics of stellar populations and star formation histories of the AVOCADO sample (Gomes et al.).
The AVOCADO dataset is thus expected to become a benchmark for comparisons with numerical simulations and high-redshift studies of dwarf galaxies.

%________________________________________________________________

\begin{acknowledgements}
The authors thank J. Lee, G. Besla and K. Johnston for useful comments and discussions.

This work was co-funded under the Marie Curie Actions of the European Commission (FP7-COFUND).
This work was partially funded by the Spanish MICINN under the Consolider-Ingenio 2010 Program grant CSD2006-00070: First Science with the GTC\,\footnote{\tt http://www.iac.es/consolider-ingenio-gtc}.
JMG is supported by grant SFRH/BPD/66958/2009 from FCT (Portugal). 
PP is supported by a Ciencia 2008 contract, funded by FCT/MCTES (Portugal) and POPH/FSE (EC).

This research has made use of VOSA, developed by the Spanish Virtual Observatory through grants AyA2008-02156 and AyA2011-24052.
The STARLIGHT project is supported by the Brazilian agencies CNPq, CAPES and FAPESP and by the France-Brazil CAPES/Cofecub program.
This research has made use of Aladin.
This research has made use of the NASA/IPAC Extragalactic Database (NED) which is operated by the Jet Propulsion Laboratory, California Institute of Technology, under contract with the National Aeronautics and Space Administration. 

Based on observations made with the NASA Galaxy Evolution Explorer.  GALEX is operated for NASA by the California Institute of Technology under NASA contract NAS5-98034.

Funding for the SDSS and SDSS-II has been provided by the Alfred P. Sloan Foundation, the Participating Institutions, the National Science Foundation, the U.S. Department of Energy, the National Aeronautics and Space Administration, the Japanese Monbukagakusho, the Max Planck Society, and the Higher Education Funding Council for England. The SDSS Web Site is http://www.sdss.org/.
The SDSS is managed by the Astrophysical Research Consortium for the Participating Institutions. The Participating Institutions are the American Museum of Natural History, Astrophysical Institute Potsdam, University of Basel, University of Cambridge, Case Western Reserve University, University of Chicago, Drexel University, Fermilab, the Institute for Advanced Study, the Japan Participation Group, Johns Hopkins University, the Joint Institute for Nuclear Astrophysics, the Kavli Institute for Particle Astrophysics and Cosmology, the Korean Scientist Group, the Chinese Academy of Sciences (LAMOST), Los Alamos National Laboratory, the Max-Planck-Institute for Astronomy (MPIA), the Max-Planck-Institute for Astrophysics (MPA), New Mexico State University, Ohio State University, University of Pittsburgh, University of Portsmouth, Princeton University, the United States Naval Observatory, and the University of Washington.

This publication makes use of data products from the Two Micron All Sky Survey, which is a joint project of the University of Massachusetts and the Infrared Processing and Analysis Center/California Institute of Technology, funded by the National Aeronautics and Space Administration and the National Science Foundation.
\end{acknowledgements}

%________________________________________________________________

\bibliographystyle{aa}
\bibliography{/Users/rsanchez/WORK/PAPERS/rsj_references.bib}

%\begin{thebibliography}{}
%  \bibitem[1997]{zheng} Zheng, W., Davidsen, A. F., Tytler, D. \& Kriss, G. A.
%     1997, preprint
%\end{thebibliography}

\end{document}